\documentclass{aa}  

\usepackage{graphicx}
\usepackage{txfonts}
\usepackage{booktabs}
\usepackage{makecell}
\usepackage{siunitx}
\usepackage[usestackEOL]{stackengine}
\begin{document}

   \title{STIX imaging I - Concept}

   \subtitle{}

     \author{Paolo Massa\inst{1} \and \inst{2}
          Gordon J. Hurford\inst{3} \and
          Anna Volpara \inst{1} \and 
          Matej Kuhar\inst{3} \and  
          Andrea Francesco Battaglia\inst{3,4} \and
          Hualin Xiao\inst{3} \and
          Diego Casadei\inst{3,5} \and
          Emma Perracchione\inst{6} \and 
          Sara Garbarino\inst{1,7} \and
          Sabrina Guastavino\inst{1} \and
          Hannah Collier\inst{3,4} \and
          Ewan C. M. Dickson\inst{8} \and	
          Daniel F. Ryan\inst{3} \and
         Shane A. Maloney\inst{9} \and 
        Frederic Schuller\inst{10} \and        
    	Alexander Warmuth\inst{10} \and
     	Anna Maria Massone\inst{1} \and    
      	Federico Benvenuto\inst{1} \and
   Michele Piana\inst{1,11} \and
      S\"am Krucker\inst{3,12}
     }

   \institute{
             MIDA, Dipartimento di Matematica, Università di Genova, via Dodecaneso 35, I-16146 Genova, Italy
             \and  
             Department of Physics \& Astronomy, Western Kentucky University, Bowling Green, KY 42101, USA
             \\\email{paolo.massa@wku.edu}
             \and      
            University of Applied Sciences and Arts Northwestern Switzerland, Bahnhofstrasse 6, 5210 Windisch, Switzerland 
  	         \and
             ETH Z\"urich, R\"amistrasse 101, 8092 Z\"urich, Switzerland 
             \and
             Cosylab Swizerland, Badenerstrasse 13, 5200 Brugg, Switzerland
             \and
             Dipartimento di Scienze Matematiche ``Giuseppe Luigi Lagrange'', Politecnico di Torino, Corso Duca degli Abruzzi, 24, 10129 Torino, Italy
            \and IRCCS Ospedale Policlinico San Martino, Largo Rossana Benzi 10, 16132 Genova, Italy
            \and 
            Institute of Physics, University of Graz, A-8010 Graz, Austria
	         \and
             School of Cosmic Physics, Dublin Institute for Advanced Studies, 31 Fitzwilliam Place, Dublin, D02 XF86, Ireland
 	         \and
            Leibniz-Institut f\"ur Astrophysik Potsdam (AIP), An der Sternwarte 16, D-14482 Potsdam, Germany
             \and
             Istituto Nazionale di Astrofisica, Osservatorio Astrofisico di Torino, Via Osservatorio 20, I-10025 Pino Torinese, Italy
             \and
             Space Sciences Laboratory, University of California, 7 Gauss Way, 94720 Berkeley, USA
}
   \date{\today}

 
  \abstract
  {}
  {To provide a schematic mathematical description of the imaging concept of the Spectrometer/Telescope for Imaging X--rays (STIX) on board Solar Orbiter. 
  The derived model is the fundamental starting point for both the interpretation of STIX data and the description of the data calibration process.
  }
  {We describe the STIX indirect imaging technique which is based on spatial modulation of the X--ray photon flux by means of tungsten grids.
  We show that each of 30 STIX imaging sub--collimators measures a complex Fourier component of the flaring X--ray source corresponding to a specific angular frequency.
  We also provide details about the count distribution model, which describes the relationship between the photon flux and the measured pixel counts.
  }
  {We define the image reconstruction problem for STIX from both visibilities and photon counts.
  We provide an overview of the algorithms implemented for the solution of the imaging problem, and a comparison of the results obtained with these different methods in the case of the SOL2022-03-31T18 flaring event.}
  {}


  \authorrunning{Massa et al.}

   \maketitle
%
\section{Introduction}

The Spectrometer/Telescope for Imaging X-rays \citep[STIX;][]{krucker2020spectrometer} is the X-ray instrument on-board Solar Orbiter \citep{2020A&A...642A...1M} and is designed to observe the X-ray radiation emitted during solar flares.
The scientific goal of STIX is to provide diagnostics of accelerated electrons and heated plasma during the energy release process in solar flares.
This task is accomplished by measuring X-ray photons emitted by thermal and non--thermal electrons through the bremsstrahlung mechanism \citep[e.g.,][]{1988psf..book.....T,krucker2008hard}.
STIX contains 32 detectors \citep{2012NIMPA.695..288M}, which measure timing and spectrum of the incident photons in the energy range $4-150$ keV. 
However, STIX is more than a simple spectrometer: thanks to an indirect imaging technique based on the spatial modulation of the incident photon flux, it is possible to reconstruct the angular distribution of the flaring X-ray source, and, hence, to perform imaging--spectroscopy analyses.

As with the Hard X-ray Telescope \citep[HXT;][]{kosugi1991hard} on board the Japanese Yohkoh mission, the Reuven Ramaty High Energy Solar Spectroscopic Imager \citep[RHESSI;][]{2002SoPh..210....3L}, and the recently launched Hard X-ray Imager \citep[HXI;][]{zhang2019hard} on board the Chinese ASO-S mission, STIX modulates the incident radiation by means of \textit{sub-collimators} (i.e., pairs of X-ray-opaque grids mounted in front of the detectors).
The modulated photon flux transmitted by the grids measures the complex values of specific Fourier components of the flaring X-ray source.
From a mathematical perspective, these complex values \textit{(visibilities)} carry the same information as that provided by antenna pair measurements in the context of radio--interferometry \citep[e.g.,][]{2017isra.book.....T}.
Hence, analogously to HXT, RHESSI, and HXI, STIX data consist of a limited number of Fourier components of the flaring source, and the image reconstruction problem for STIX consists in solving the ill-posed (see Sect. \ref{section:imging_methods}) inverse problem of retrieving the image of the X-ray source from a set of corresponding visibilities \citep{pianabook}. 

Despite these similarities, STIX, HXT, RHESSI, and HXI modulate the X-ray radiation and measure visibilities in different ways.
HXT and HXI measure the real and the imaginary part of each Fourier component using a pair of sub--collimators, whose grids have the same orientation and pitch. 
The two sub--collimators of the same pair are different in the relative position of the front and rear grid, which is shifted by a quarter of a pitch.
This causes the corresponding transmission functions to be shifted in phase by 90 degrees. 
Hence, while the transmission function of the sub--collimator measuring the real part of the visibility is related to a cosine function, the transmission function of the sub--collimator measuring the imaginary part is related to a sine function with the same period and orientation.
RHESSI temporally modulates the X-ray radiation by means of nine Rotating Modulation Collimators \citep[RMC;][]{2002SoPh..210...61H}, in which the front and the rear grid have identical parallel slits. 
The modulation effect of the collimators is obtained by means of the instrument rotation around its optical axis, which has a period of \(\sim \)4 s. 
From the modulated count-rate profiles recorded by each detector during a rotation, it is possible to retrieve the values of a few hundred visibilities, corresponding to angular frequencies located on nine concentric circles in the spatial frequency $(u,v)$--plane.
Each circle corresponds to a different angular resolution and is sampled by a specific collimator.
On the other hand, the X-ray photon flux transmitted through the grid pair of each STIX sub--collimator creates a spatial Moir\'e pattern \citep[e.g.,][]{prince1988gamma,hurford2013x} on the corresponding detector surface.  Measurements of this pattern performed by a pixelated detector allow the determination of amplitude and phase of the corresponding STIX visibility.

In this study, which is the first of a series of three papers dealing with STIX imaging, we provide a mathematical description of the imaging concept of the STIX instrument.
Specifically, assuming \textit{ideal} sub--collimators, we derive the analytical expressions of 1) the transmission function of the sub--collimators;  2) the Moir\'e patterns created by the photon flux transmitted through the front and rear grids; 3) the angular frequencies sampled by the sub--collimators; and 4) the relationship between the number of X-ray counts recorded by the detectors and the values of the visibilities' amplitude and phase.
Further, we describe the STIX count distribution model, which is the linear map connecting the photon flux directly to the count measurements performed by STIX detectors \citep{massa2019count, Siarkowski}.
We derive general results concerning the description of the Moir\'e pattern technique and its use for indirect imaging of solar flares \citep[e.g.,][]{prince1988gamma,hurford2013x} for the specific case of STIX and we extend previous works by \cite{gietal15}, \cite{massa2019count}, and \cite{Siarkowski}, by deriving a general and exhaustive description of the visibility 
formation process and of the count distribution model.

The main merits of this paper are that 1) the model it describes can be used to define and address the STIX image reconstruction problem; and 2) it represents the starting point for the description of the STIX data calibration process. In fact, in the following two papers of the series we will discuss
\begin{itemize}
\item The \textit{on--ground} data calibration process, including both corrections that are not considered in the present paper (e.g., energy dependence of the grid response, internal shadowing of the grids, partial transparency of the grids for photons passing through the corner of a slat), and the characterization of the parameters of the STIX grids that has been performed before the instrument was assembled.
\item The \textit{in--flight} self--calibration techniques based on built-in data redundancy, which has been exploited after the launch of the instrument in February 2020.  
\end{itemize}

The plan of this paper is as follows. In Sect. \ref{section:hardware_parameters} we describe the STIX imager hardware and the parameters characterizing the grids and the detectors. Section \ref{section:assumptions} deals with the assumptions we made for the mathematical description of the imaging concept. 
Section \ref{section:data-formation} contains the key result of this paper, which is the description of the visibility formation process and of the count distribution model.
In Sect. \ref{section:def_fov} we describe the different Field-of-Views (FOVs) of the STIX instrument, while in  Sect. \ref{section:imging_methods} we provide a brief summary of the image reconstruction problem for STIX and of the imaging techniques implemented so far for its solution, as illustrated by an analysis of the SOL2022-03-31T18 flare.
Finally, our conclusions are offered in Sect. \ref{section:conclusions}.

\begin{figure*}[t]
\centering

\includegraphics[width=0.8\textwidth]{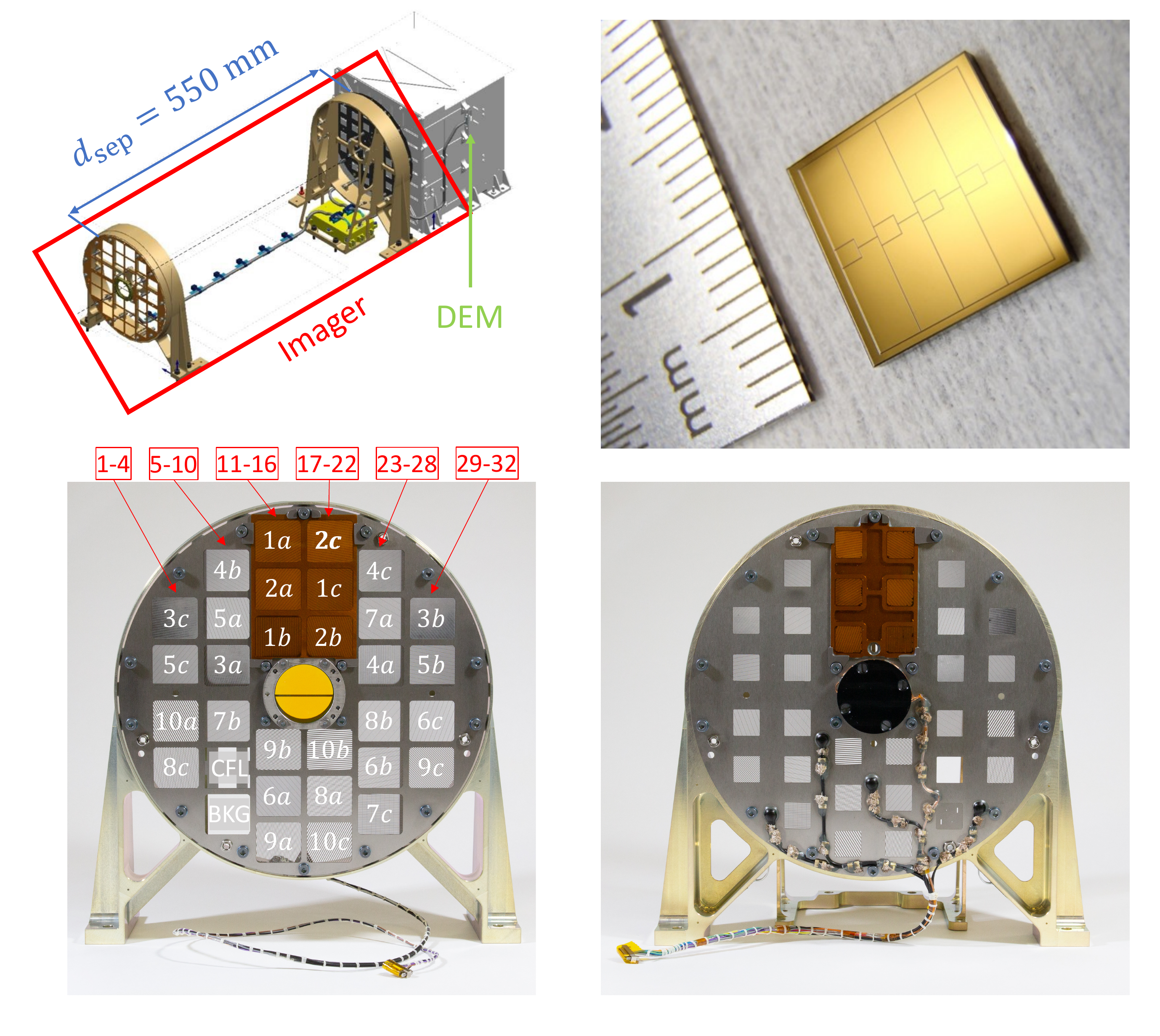}

\caption{\textit{Top left panel:} schematic of the STIX instrument showing the imager and the DEM.
The separation between the front and the rear grids is indicated with a blue arrow.
\textit{Top right panel:} STIX CdTe detector.
\textit{Bottom panels:} pictures of the front and the rear grids of the STIX flight spare.
In the bottom-left panel, the sub--collimator labels are reported within the corresponding windows, while the sub--collimator numbers are reported in the red boxes. 
Since the order of the sub--collimators decreases from top to bottom in each column, the first and the last sub--collimator number of each column is reported in the corresponding box.}
\label{fig:scheme-imager}
\end{figure*}

\section{The STIX hardware parameters}\label{section:hardware_parameters}

The STIX imager is composed of a front and a rear tungsten grid, separated by a nominal distance $d_{\text{sep}} = 550$ mm (see Fig. \ref{fig:scheme-imager}, top left panel).
Each grid contains 32 \textit{windows} (i.e., sub-areas consisting of parallel slats and corresponding slits), which are shown in the bottom panels of Fig. \ref{fig:scheme-imager}. 
The front and the rear grid windows have nominal dimension 22 mm $\times$ 20 mm and 13 mm $\times$ 13 mm, respectively.
The centers of the windows are nominally aligned with the center of the corresponding Cadmium-Telluride (CdTe) detectors (\cite{2012NIMPA.695..288M}; see the top right panel of Fig. \ref{fig:scheme-imager}), which are mounted in the Detector Electronic Module (DEM, see Fig. \ref{fig:scheme-imager}, top left panel), at a nominal distance
$d_{\text{det}} = 47$ mm from the rear grid. 
In the following we will denote as \textit{sub-collimator} the unit consisting of the front window, the rear window and the detector. Moreover, with a slight abuse of language, we will sometimes refer to the windows as grids.

The STIX sub--collimators are denoted by a number between 1 and 32. 
Among them, 30 are used for imaging purposes, while the remaining two (sub--collimators 9 and 10) are the Coarse Flare Locator (CFL) and the Background Monitor (BKG).
The CFL provides an estimate of the flare location with an uncertainty of a couple of arcmins, while the BKG measures the background radiation or, during intense events, the low energy flux (see Figs. 8 and 9 in \citet{krucker2020spectrometer} for a schematic picture of the CFL and BKG apertures).
The 30 imaging sub--collimators are also denoted by a label, which consists of a number, between 1 and 10, and a letter ($a$, $b$ or $c$): the number indicates the resolution of the sub-collimator (detectors with finest resolution are those associated with lowest numbers), while the letter refers to the orientation of the grids.
The number and the label associated with each sub--collimator are reported in the bottom--left panel of Fig. \ref{fig:scheme-imager}.

\begin{figure*}[t]
\centering
\includegraphics[width=0.8\linewidth]{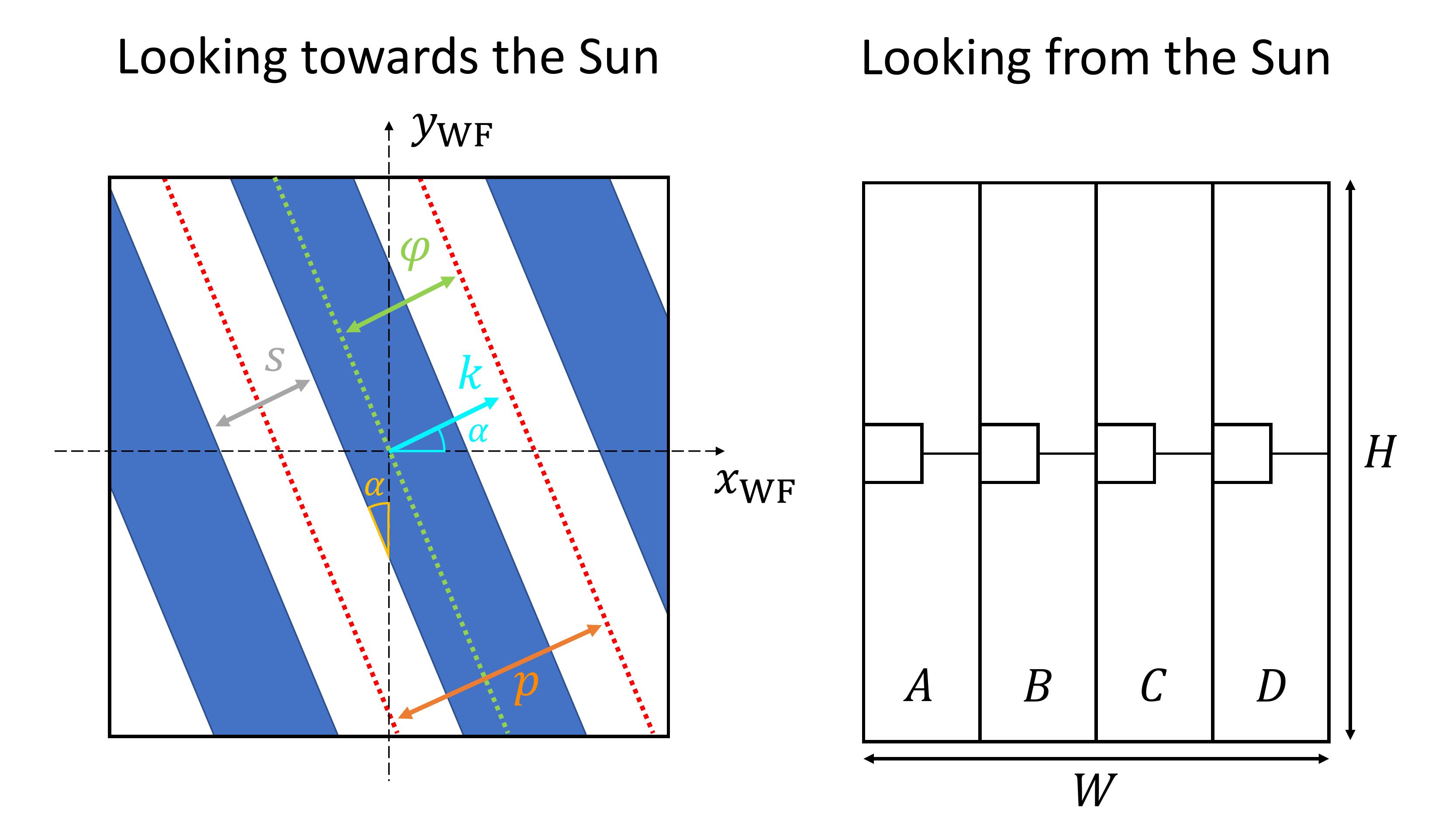}
\caption{\textit{Left panel:} schematic of a grid window and of its characterizing parameters: pitch $p$, slit width $s$, phase $\varphi$, orientation angle $\alpha$,  and wave vector $\mathbf{k}$. \textit{Right panel:} schematic of a STIX detector, showing the division into vertical stripes and pixels. The detector width and height are denoted by $W$ and $H$, respectively, while $A$, $B$, $C$, and $D$ indicate the number of X-ray counts recorded by the pixels in the corresponding vertical stripe (from left to right, respectively).}
\label{fig:scheme-grid}
\end{figure*}

Let us consider a window of an imaging sub--collimator (either the front or the rear one, as seen from the detector side looking towards the Sun) and let us fix a coordinate system located at the center of the window itself, with axes parallel to the window edges\footnote{Adjacent edges of the windows and of the detectors are nominally orthogonal.}.
We denote this coordinate system as \textit{Window Frame} (WF).
The window is characterized by the following parameters (see Fig. \ref{fig:scheme-grid}):
\begin{itemize}
\item slit width $s$;
\item period $p$ (also named \textit{pitch}), which is the sum of the widths of a slit and a slat;
\item orientation angle $\alpha$, measured counterclockwise from the positive $y$ semi-axis, and wave vector $\mathbf{k} = (\cos(\alpha), \sin(\alpha))$, which is perpendicular to the slits;
\item phase $\varphi$, which is the distance between the window center and the first slit center encountered in the direction determined by $\mathbf{k}$.
\end{itemize}
Each window is also characterized by a nominal thickness $t=0.4$ mm, which is assumed to be zero in this paper (see Sect. \ref{section:assumptions} below).
In practice, since the grid thickness is not negligible, X--ray photons produced by off--axis sources may be stopped by the internal sides of the window slats. 
This \textit{internal shadowing} effect results in a reduction of the effective slit width for off--axis sources and has to be taken into account during the data calibration process, which will be discussed in the second paper of the series.

By design, the values of the parameters of the front and the rear window of the same sub--collimator are slightly different. 
Hence, in the following we will use the subscripts $f$ and $r$ to refer to parameters of the front and of the rear grid, respectively.
A fundamental relationship that is satisfied by the parameters of the front and of the rear window of each STIX sub--collimator is \citep{gietal15}
\begin{equation}\label{eq: wave vector relations}
\frac{1}{p_f} \mathbf{k}_f - \frac{1}{p_r} \mathbf{k}_r  =  \left( \pm \frac{1}{W}, 0 \right) ~,
\end{equation}
where, at the right--hand side, $W$ denotes the width of the corresponding detector and the sign is plus if $p_r \cos(\alpha_f) > p_f \cos(\alpha_r)$, and minus otherwise. 
The sign factor $\pm 1$ is called \emph{m} factor.
We will show in Sect. \ref{section:data-formation} that the transmitted flux creates a bi--dimensional sinusoidal distribution on the detector surface, and that the vector defined at the right hand side of \eqref{eq: wave vector relations} determines period and orientation of such a wave.
The $m$ factor indicates in which direction the distribution moves depending on the location of the X--ray source. 
For self--calibration purposes, which will be discussed in the third paper of the series, the sub--collimators are designed so that half of them has $m$ factor equal to 1 (and the remaining half equal to -1).

In Table \ref{tab:nominal_values} we report the nominal values of the parameters characterizing each sub--collimator: the coordinates of the center of the windows, the slit width ($s$), pitch ($p$), and orientation angle ($\alpha$) of the windows, and the $m$ factor.
The coordinates of the window centers are measured with respect to a coordinate system fixed in the center of the grid (either the front or the rear one) and with axes parallel to the window edges \citep[see Chapter 2 of][]{kuhar2018through}. 
Note that the nominal slit--to--pitch ratio ($s/p$) is always slightly larger than 0.5. 
The slit width is chosen as \(\sim\)3 $\mu$m larger than half of the average pitch of the front and rear windows.
This is done to compensate for the reduction of the effective slit width for off--axis sources due to the internal shadowing effect.
Thanks to this choice, the slit--to--pitch ratio is on average (with respect to the possible locations of the X--ray source) closer to 0.5, and this guarantees a higher signal-to-noise ratio (S/N) for the measured visibility values.

We point out that the window phases have no nominal values and, hence, are not reported in Table \ref{tab:nominal_values}.
Indeed, as we will show in Sect. \ref{section:data-formation}, the values of the window phases do not affect the location of the frequencies sampled by STIX; instead, they have to be taken into account only in the computation of the visibility phases. 
Hence, window phases are not design parameters and their values were determined only after the fabrication of the grids, during the grid characterization process that has been performed on ground using optical and X-ray measurements collected at the
Paul Scherrer Institute (PSI), Switzerland. A more detailed description of the grid characterization process will be provided in a follow-up paper, together with the values of the measured grid
parameters.

The windows of the six detectors with finest resolution (labelled with $1$-$a,b,c$ and $2$-$a,b,c$) have a different structure.
Manufacturing limitations made it necessary to build them as a superimposition of two layers (in the case of sub-collimators $2$-$a,b,c$) or three layers (in the case of sub-collimators $1$-$a,b,c$).
The thickness of each layer is 1/2 or 1/3 of the nominal grid thickness. 
Each layer grid has a period which is twice or three times larger than the nominal pitch of the overall grid, and wide slits.
The effective nominal pitches of 0.054 mm (for sub-collimators $2$-$a,b,c$) and 0.038 mm (for sub-collimators $1$-$a,b,c$) are then obtained by mounting the different layer grids with a difference in phase of a half or a third of their pitch, respectively.

\begin{figure}[ht]
\centering
\includegraphics[width=0.8\columnwidth]{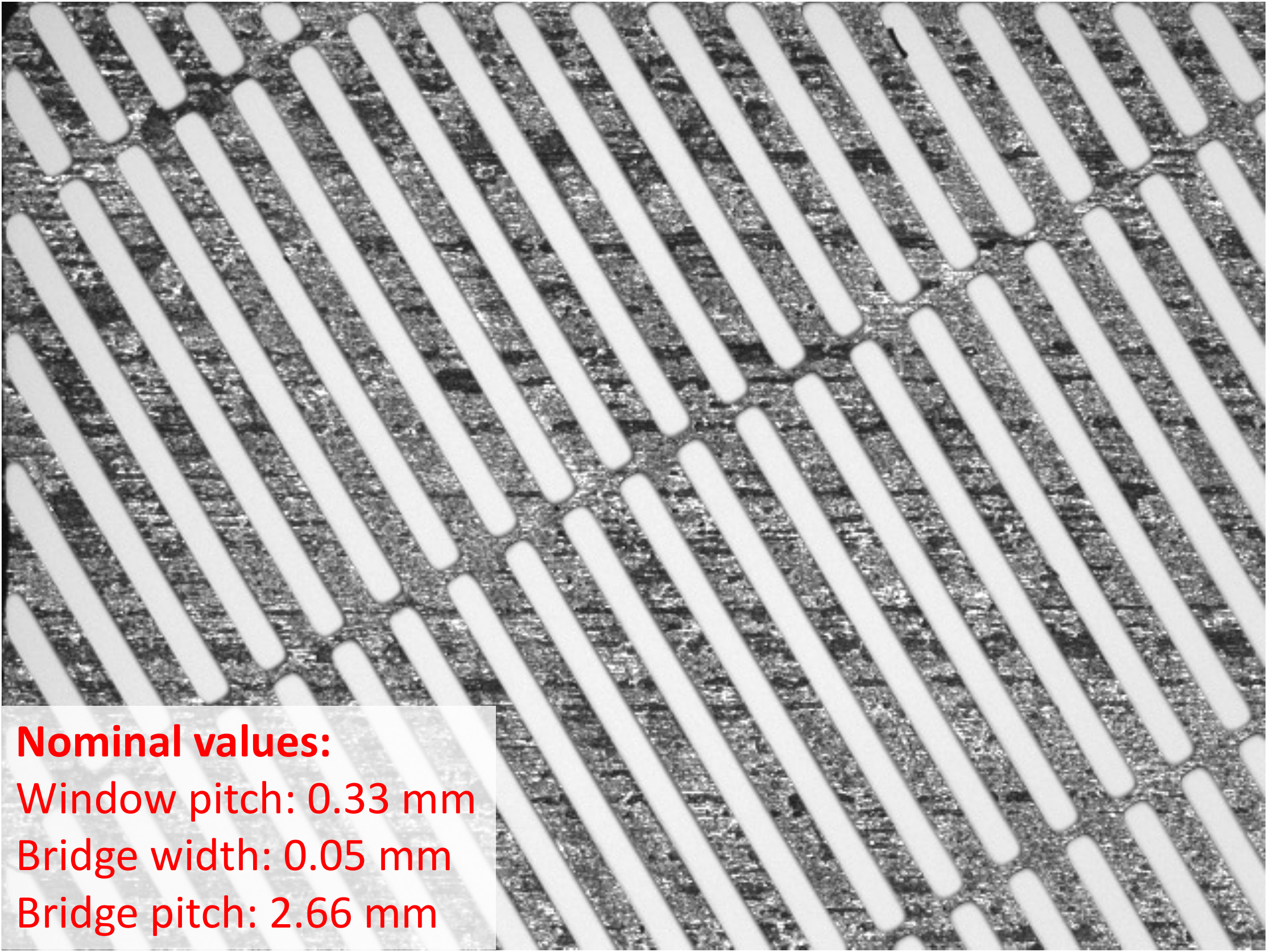}
\caption{Close-up of the rear window of sub--collimator 8 ($7b$) showing bridges oriented in the bottom left--top right direction.
The nominal values of the window pitch, bridge width, and bridge pitch are reported in the bottom--left corner.}
\label{fig:bridges}
\end{figure}

The windows of the sub--collimators labelled with $1$-$a,b,c$ through $8$-$a,b,c$ present narrow tungsten stripes connecting the slats, named \textit{bridges}, which are added for increasing the robustness of the grids. 
As shown in Fig. \ref{fig:bridges}, bridges are perpendicular to the slats and have a wide period.
The presence of the bridges does not affect the sub--collimator modulation of the X-ray flux. 
The only effect of the bridges is to slightly decrease the grids' transmission, and appropriate corrections accounting for this effect are performed during the data calibration process, as we will discuss in the second paper of the series.

\begin{table*}[t]
    \centering
    \resizebox{\linewidth}{!}{\begin{tabular}{cccccccccc}
    \toprule
    \Longunderstack{\textbf{Sub--coll.} \\ \textbf{label}}
    &\Longunderstack{\textbf{Sub--coll.} \\ \textbf{number}}
    &\Longunderstack{\textbf{Center X} \\ \textbf{[mm]}} &\Longunderstack{\textbf{Center Y} \\ \textbf{[mm]}} 
    &\Longunderstack{\textbf{Slit width}\\\textbf{[mm]}} 
    & \Longunderstack{\textbf{Pitch front} \\ \textbf{[mm]}} &\Longunderstack{\textbf{Orientation front}\\\textbf{[deg]}}
    & \Longunderstack{\textbf{Pitch rear} \\ \textbf{[mm]}} &\Longunderstack{\textbf{Orientation rear}\\\textbf{[deg]}}
    &\Longunderstack{\textbf{$m$}\\\textbf{factor}}\\
    \midrule
    10a &3 &-62.5 &-13.5 &0.479 &0.909644 &151.481 &0.999045 &148.374 &-1\\
10b &20 &12.5 &-27.5 &0.479 &0.951208 &86.902 &0.951208 &93.098 &1\\
10c &22 &12.5 &-73.5 &0.479 &0.909644 &28.519 &0.999045 &31.626 &1\\
9a &16 &-12.5 &-73.5 &0.336 &0.641967 &170.363 &0.691656 &169.609 &-1\\
9b &14 &-12.5 &-27.5 &0.336 &0.674193 &107.937 &0.656988 &112.010 &1\\
9c &32 &62.5 &-36.5 &0.336 &0.682197 &51.702 &0.649830 &48.379 &-1\\
8a &21 &12.5 &-50.5 &0.236 &0.453678 &9.744 &0.477944 &10.270 &1\\
8b &26 &37.5 &-13.5 &0.236 &0.457628 &131.141 &0.473450 &128.819 &-1\\
8c &4 &-62.5 &-36.5 &0.236 &0.461187 &68.589 &0.469602 &71.437 &1\\
7a &24 &37.5 &36.5 &0.166 &0.320259 &29.479 &0.330680 &30.538 &1\\
7b &8 &-37.5 &-13.5 &0.166 &0.320259 &150.521 &0.330680 &149.462 &-1\\
7c &28 &37.5 &-59.5 &0.166 &0.325344 &91.059 &0.325344 &88.941 &-1\\
6a &15 &-12.5 &-50.5 &0.117 &0.229395 &50.572 &0.225614 &49.437 &-1\\
6b &27 &37.5 &-36.5 &0.117 &0.230433 &169.870 &0.224640 &170.127 &1\\
6c &31 &62.5 &-13.5 &0.117 &0.228493 &109.301 &0.226482 &110.693 &1\\
5a &6 &-37.5 &36.5 &0.083 &0.158505 &69.515 &0.159487 &70.488 &1\\
5b &30 &62.5 &13.5 &0.083 &0.160427 &10.091 &0.157598 &9.911 &-1\\
5c &2 &-62.5 &13.5 &0.083 &0.158078 &130.394 &0.159925 &129.601 &-1\\
4a &25 &37.5 &13.5 &0.059 &0.111198 &89.638 &0.111198 &90.362 &1\\
4b &5 &-37.5 &59.5 &0.059 &0.110594 &29.820 &0.111811 &30.182 &1\\
4c &23 &37.5 &59.5 &0.059 &0.110594 &150.180 &0.111811 &149.818 &-1\\
3a &7 &-37.5 &13.5 &0.042 &0.077582 &110.237 &0.077817 &109.762 &-1\\
3b &29 &62.5 &36.5 &0.042 &0.077921 &50.194 &0.077480 &49.807 &-1\\
3c &1 &-62.5 &36.5 &0.042 &0.078039 &169.956 &0.077364 &170.044 &1\\
2a &12 &-12.5 &50.5 &0.030 &0.054408 &129.864 &0.054192 &130.135 &1\\
2b &19 &12.5 &27.5 &0.030 &0.054357 &70.166 &0.054243 &69.834 &-1\\
2c &17 &12.5 &73.5 &0.030 &0.054136 &9.969 &0.054465 &10.031 &1\\
1a &11 &-12.5 &73.5 &0.022 &0.037929 &150.062 &0.038071 &149.938 &-1\\
1b &13 &-12.5 &27.5 &0.022 &0.038000 &90.124 &0.038000 &89.876 &-1\\
1c &18 &12.5 &50.5 &0.022 &0.037929 &29.938 &0.038071 &30.062 &1\\
    \bottomrule
    \end{tabular}}
\caption{Nominal values of the parameters characterizing the STIX imaging sub--collimators.
\textit{From left to right}: sub--collimator label, sub--collimator number, coordinates of the window center, slit width, pitch of the front window, orientation angle of the front window, pitch of the rear window, orientation angle of the rear window, and $m$ factor.
The values of the center coordinates and of the slit width are equal for both the front and the rear windows of the same sub--collimator.
The parameters of sub--collimators 1-$a$, $b$, $c$ and 2-$a$, $b$, $c$ refer to the windows as a whole, without considering the division into different layers (see the text for more details), while the parameters of sub--collimators 9 and 10 are not reported as they refer to CFL and BKG, respectively.}
\label{tab:nominal_values}
\end{table*}

STIX detectors have nominal width $W=8.8$ mm and height $H=9.2$ mm (see the right panel of Fig. \ref{fig:scheme-grid}). 
Each detector is vertically partitioned into four identical stripes and each stripe consists of three units, named \textit{pixels}. 
Nominally, the \textit{top} and the \textit{bottom} pixels have the same area equal to 0.096 mm$^2$, while the area of the central \textit{small} pixel is $0.01$ mm$^2$.
Each pixel independently measures the number of incident X-ray photons. In the following, we will denote by $A$, $B$, $C$, and $D$, the counts recorded by the pixels in the four stripes (from left to right, looking at the detector from the Sun side).
As we will describe in Sect. \ref{section:data-formation}, the detector partition into four vertical stripes is crucial for imaging purposes.
On the other hand, the reason for the additional partitioning of each stripe into three pixels is twofold.
First, during large flares, top and/or bottom row pixels can be disabled by the Rate Control Regime (RCR) algorithm in order to reduce the incident area and, consequently, increase the detector live time \citep{krucker2020spectrometer}. 
Second, the three rows of pixels (top, small and bottom) provide redundant information that can be exploited for self-calibration purposes, as we will discuss in the third paper of the series.

\section{Assumptions and definitions}\label{section:assumptions}

\begin{figure*}[ht]
    \centering
    \includegraphics[width=\textwidth]{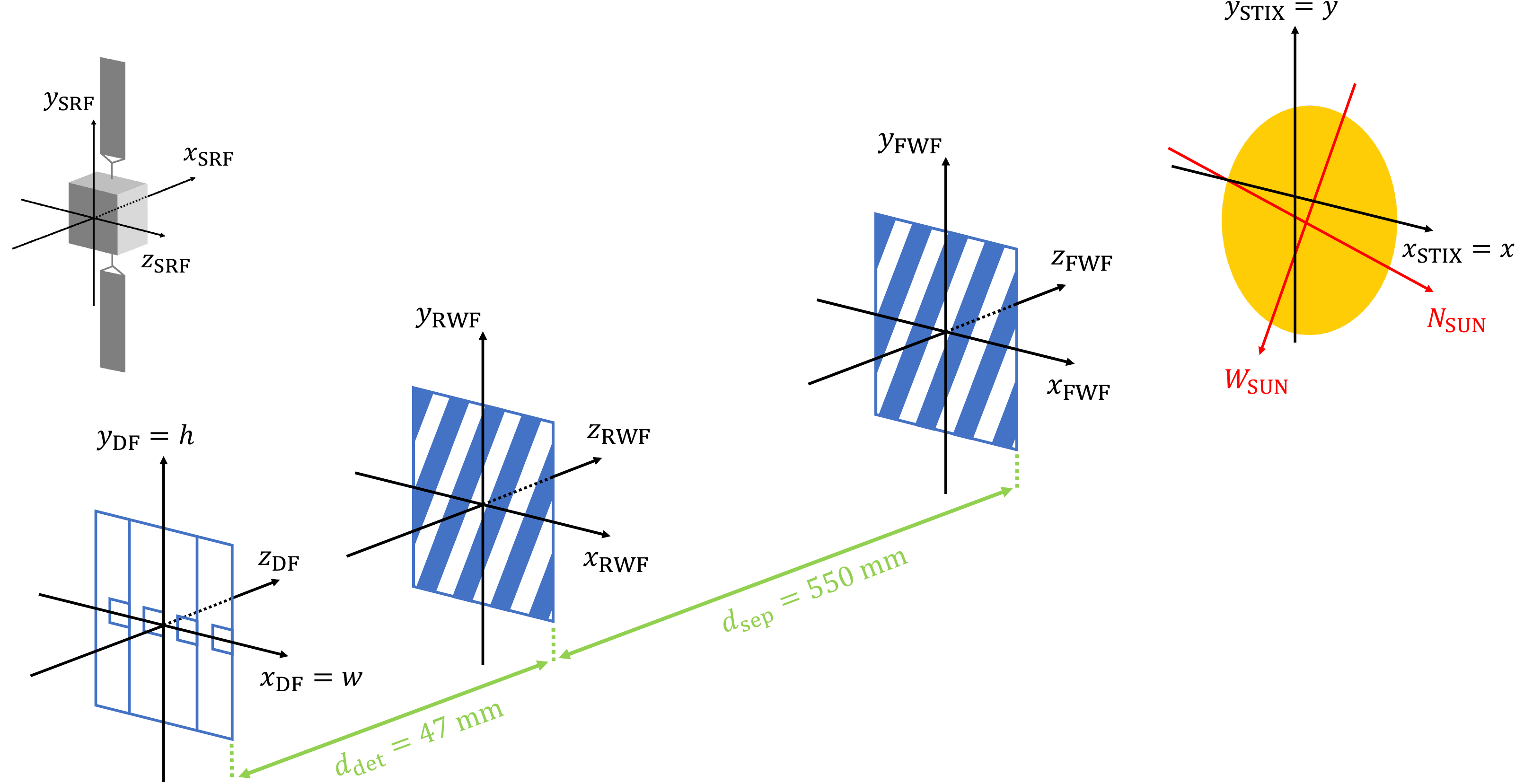}
    \caption{Schematic representing the DF reference frame defined on the detector surface, the RWF defined on the rear window, the FWF defined on the front window and the angular $(x_{\mathrm{STIX}},y_{\mathrm{STIX}})$ coordinates defined on the solar disk.
    The North and West axes of the Helioprojective-Cartesian (HPC) reference frame are reported in red on the solar disk.
    A schematic picture of Solar Orbiter is reported in the top--left corner in order to show the relationship between the DF, RWF, and FWF reference frames and the Spacecraft Reference Frame (SRF).
    The green arrows denote the nominal distances between the front and the rear window, and between the rear window and the detector.
    Note that the DF reference frame and the HPC reference frame are not centered in the same point and are not exactly 90 deg rotated in order to show the typical situation in which the instrument is slightly off-pointing and the spacecraft roll-angle is not negligible.}
    \label{fig:reference_frames}
    \end{figure*}

Before starting the description of the mathematical model of STIX data, we need to introduce a few assumptions and definitions, which are valid throughout the paper.
\begin{itemize}
    \item In the following, we will consider a single sub--collimator; the results we will derive are completely general and apply to each STIX sub--collimator.
    \item The considered sub--collimator is assumed to be \textit{ideal}.
    Specifically, we assume that its windows consist of a single layer and have parallel slats
    perfectly opaque to X-ray radiation.
    Also, we assume that the windows have negligible thickness and that bridges are not present. 
    We will show in the second paper of the series how to generalize the results of this work to the case of a \textit{real} sub--collimator by accounting for several issues during the data calibration process, including energy dependence of the grid response, non--negligible grid thickness, and presence of bridges.
    
    \item We define a reference system, named \textit{Detector Frame} (DF), on the Sun side of the detector surface. 
    The axes $x_{\text{DF}}$ and $y_{\text{DF}}$ are parallel to the horizontal and vertical detector edges, respectively; the remaining $z_{\text{DF}}$ points towards the Sun (see Fig. \ref{fig:reference_frames}). Analogously, we consider a reference frame on both the front and the rear windows, named \textit{Front Window Frame} (FWF) and \textit{Rear Window Frame} (RWF), which have $x$ and $y$ axes parallel to the window edges and $z$ axes pointing towards the Sun (see Fig. \ref{fig:reference_frames}). 
    We assume that the axes $z_{\text{DF}}$, $z_{FWF}$, and $z_{RWF}$ are parallel, and that the centers of the detector and of the front and rear windows are aligned.
    Points on the Sun surface are assumed to be far away in the $z$ direction of the three reference frames, so that for them we can assume $z_{\text{DF}}=z_{\text{FWF}}=z_{\text{RWF}}$.
    Moreover, will assume that the Sun is a flat disk as seen from STIX, and that points on that disk are described by angular coordinates $(x_{\mathrm{STIX}},y_{\mathrm{STIX}})$, where 
    \begin{equation}
    \begin{split}
    & x_{\mathrm{STIX}} = \tan^{-1}\left(\frac{x_{\mathrm{DF}}}{z_{\mathrm{DF}}} \right) = \tan^{-1}\left(\frac{x_{\text{FWF}}}{z_{\text{FWF}}} \right) = \tan^{-1}\left(\frac{x_{\text{RWF}}}{z_{\text{RWF}}} \right) ~, \\
    & y_{\mathrm{STIX}} = \tan^{-1}\left(\frac{y_{\mathrm{DF}}}{z_{\mathrm{DF}}} \right) = \tan^{-1}\left(\frac{y_{\text{FWF}}}{z_{\text{FWF}}} \right) = \tan^{-1}\left(\frac{y_{\text{RWF}}}{z_{\text{RWF}}} \right) ~.
    \end{split}
    \end{equation}
    Since STIX dimensions are negligible compared to the Sun dimension, we can assume that the $(x_{\mathrm{STIX}},y_{\mathrm{STIX}})$ reference frame is the same for every sub--collimator. This reference frame is the one adopted for image reconstruction purposes (i.e., the angular coordinates of the maps reconstructed from STIX data are expressed with respect to this frame). As shown in Fig. \ref{fig:reference_frames}, STIX is mounted on a side of the Solar Orbiter spacecraft and, hence, it observes the Sun as 90--degree--rotated.
    Assuming that the spacecraft roll angle is equal to zero and that the instrument points exactly towards the center of the solar disk, the relation between the $x_{\mathrm{STIX}}$ and $y_{\mathrm{STIX}}$ axes and the solar North ($N_{\text{SUN}}$) and West ($W_{\text{SUN}}$) axes of the Helioprojective-Cartesian coordinate frame \citep[HPC;][]{2006A&A...449..791T} is given by
    \begin{equation}\label{eq:eq_rot_stix_sun}
    \begin{split}
    x_{\mathrm{STIX}} &= N_{\text{SUN}}\\
    y_{\mathrm{STIX}} &= -W_{\text{SUN}}
    \end{split}~.
    \end{equation}
   Thanks to the short--term stability of the spacecraft pointing, flare image reconstruction can proceed without reference to an aspect solution. However, absolute positioning of the final image must take this into account.  
   For pitch and yaw this is done by using the STIX Aspect System \citep[SAS;][]{2020SoPh..295...90W},
   whose optical elements are integral parts of the front and rear grids to determine the location of the solar limbs relative to the X--ray grids themselves.   Roll aspect relies on the as-flown spacecraft aspect solution. 
   Although the typical accuracy of \(\sim\)10 arcsec currently achieved falls short of the design goal of 4 arcsec, ongoing improvements to the aspect analysis should reach this goal in the near future and can be retroactively applied to previously analyzed events.   As an interim measure, fine tuning of the final position of the images can make use of observations at other wavelengths.

    More details concerning the reference frames involved and the operations to perform for accurate image placement will be given in the second paper of the series.
    Finally, for simplifying the notations, points $(x_{\text{DF}}, y_{\text{DF}}, 0)$ on the detector surface will be denoted as $(w,h)$ and the angular coordinates $(x_{\mathrm{STIX}}, y_{\mathrm{STIX}})$ describing points on the solar disk will be denoted as $(x,y)$.

    \item The transmission function of the considered \textit{ideal} windows will be modelled as a square wave function and it will be approximated by truncating its Fourier expansion at the fundamental harmonic.
    This approximation will allow the determination of the analytic expression of the sub--collimator transmission function and of the Moir\'e pattern created by the transmitted flux.
    The detected flux due to the contribution of the higher harmonics is assumed to be negligible\footnote{Although even harmonics in the Fourier expansion of the window transmission functions can be non--zero, the detector pixelization fully suppresses any affect on the measurement of the fundamental (Eq. \eqref{eq: counts ABCD 1}).
    The measurement of the third harmonic of the front and the rear window transmission functions result in a visibility amplitude at the third harmonic detected flux which is less than 4$\%$ of the fundamental The contribution of higher odd harmonics is negligible.} and will therefore be neglected in the model derived in this paper.
   
\end{itemize}

\begin{figure*}[ht]
\centering
\includegraphics[width=\textwidth]{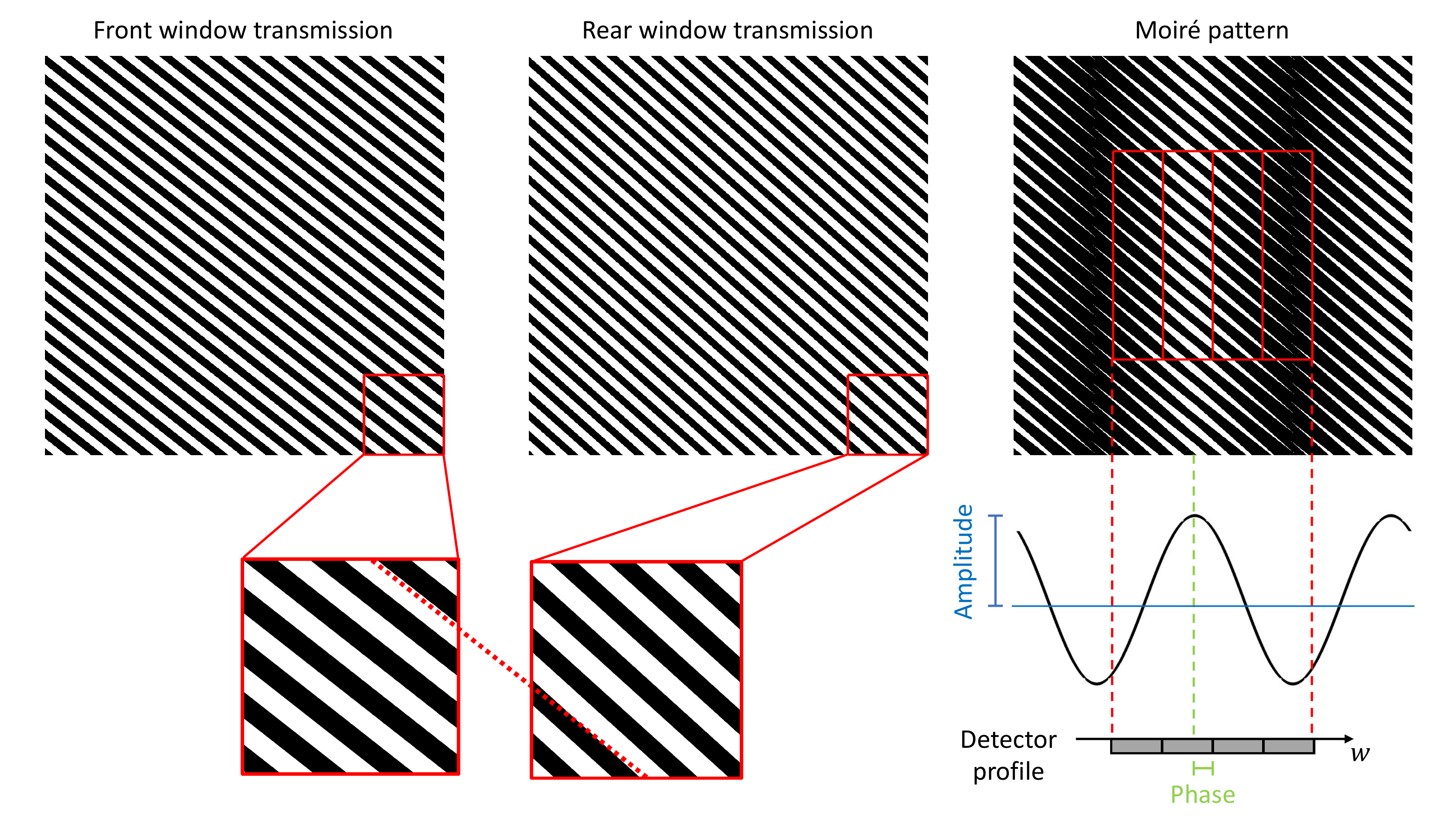}\\
\caption{\textit{Top left and middle panels:} simulation of the flux transmitted through the front and rear windows of sub-collimator 32 (9$c$), respectively.
The simulated flux corresponds to a point--like X-ray source located in $(x,y)=(0,0)$. 
\textit{Top right panel:} Moir\'e pattern created by the superimposition of the flux transmitted through the windows.
The edges of the detector vertical stripes are reported in red.
The pattern is plotted as seen looking towards the Sun from behind the detector.
\textit{Bottom left and middle panels:} close--up of the flux transmitted through the front and rear windows, respectively.
A red dashed line is plotted as a reference to show that the orientations of the front and  rear window slits are slightly different.
\textit{Bottom right panel:} profile of the Moir\'e pattern (approximated as a sinusoidal wave) and of the detector along the $w$-axis. Phase and amplitude of the pattern are indicated in green and in blue, respectively.}
\label{fig:moire-pattern}
\end{figure*}

Finally, we will denote by $\phi(x,y)$ the function representing the number of X-ray photons emitted per unit of time, energy and area from a point $(x,y)$ on the solar disk, and incident per unit of area on the telescope. 
The units of $\phi$ are counts arcsec$^{-2}$ cm$^{-2}$ s$^{-1}$ keV$^{-1}$; however, throughout the paper $\phi$ will be considered as already integrated over appropriate time and energy intervals (its units being then counts arcsec$^{-2}$ cm$^{-2}$), since this integration is not relevant for the description of the visibility formation process and of the count distribution model.

\section{Mathematical model of STIX data}\label{section:data-formation}

In this section we describe the relationship between the counts recorded by the detector pixels and the complex value of a specific Fourier component of the flaring X-ray source.
Then, we will provide details about the count distribution model, which directly links the distribution of the X-ray source to the recorded pixel counts.

\subsection{Visibility formation process}\label{section:visibility}

The \textit{visibility} corresponding to an angular frequency $(u,v)$ is defined as the Fourier component \citep{gietal15}
\begin{equation}
V(u,v) = \iint \phi(x,y) \exp\left(2\pi i (xu+yv) \right) \,dx\,dy ~.     
\end{equation}
In the following we will show that each STIX sub--collimator measures a specific visibility. 
In particular, amplitude and phase of the complex value $V(u,v)$ are determined by the number of counts $A$, $B$, $C$, and $D$ measured by the detector pixels, while the frequency $(u,v)$ is determined by pitch and orientation of the front and rear window. 
By design, pitch and orientation of the front window are slightly different from those of the rear window (see Table \ref{tab:nominal_values}). 
Thanks to this characteristic, the X-ray flux transmitted through the sub--collimator windows gives rise to a \textit{Moir\'e pattern} \citep{prince1988gamma,hurford2013x} on the detector surface. 
Figure \ref{fig:moire-pattern} shows a simulation of the Moir\'e pattern created by the flux transmitted through the windows of sub--collimator 32 (9$c$) and corresponding to an ideal point--like X-ray source\footnote{An ideal point--like source is defined as $\phi(x,y) = \delta_{(x_c,y_c)}(x,y)$, where $\delta_{(x_c,y_c)}$ is the Dirac-delta centered at $(x_c,y_c)$.} located in $(x,y)=(0,0)$.
In the following we will show that the Moir\'e pattern can be approximated as a bi--dimensional sinusoidal function with period equal to the detector width $W$ and direction parallel to the $w$-axis of the detector reference frame. 
Further, we will demonstrate that the pattern's amplitude, which is defined as half the difference between the maximum and the minimum value of the sinusoidal wave, is proportional to the amplitude of a specific complex visibility and that the pattern's phase, which represents the distance between the location of the peak of the sinusoidal wave and the center of the detector, corresponds to the phase of the visibility (see the bottom--right panel of Fig. \ref{fig:moire-pattern}).

\begin{figure*}[ht]
\centering
\includegraphics[width=\textwidth]{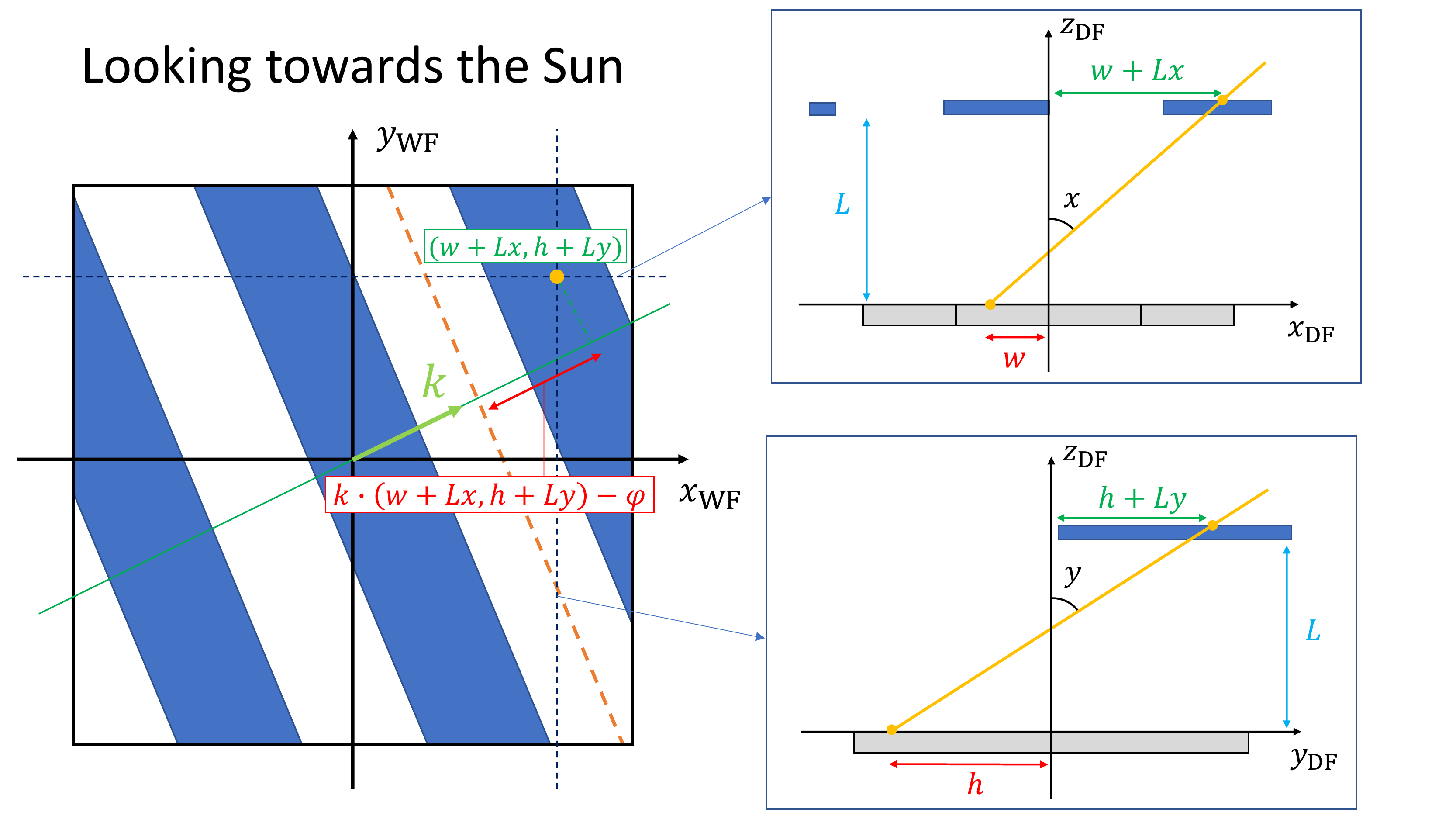}\\
\caption{Schematic of the photon trajectory (yellow line) incident in the point $(w,h)$ on the detector surface, with incident angles $x$ and $y$ with respect to the $x_{\text{DF}}$ and $y_{\text{DF}}$ axes.
\textit{Left panel:} intersection point between the trajectory and the $(x_{\text{WF}},y_{\text{WF}})$-plane of the Window Frame (either the front or the rear one). 
The red arrow denotes the distance between the intersection point and the reference slit center, which is plotted with an orange dashed line. 
The wave vector and the phase of the window are denoted by $\mathbf{k}$ and $\varphi$, respectively (see Sect. \ref{section:hardware_parameters}), while $L$ denotes the distance between the window and the detector.
\textit{Top and bottom right panels:} profiles of the trajectory along the dashed lines parallel to the $x_{\text{WF}}$ and $y_{\text{WF}}$ axes that are shown in the left panel. 
We note that, in this specific example, the values of $w$ and $h$ are negative, since they have been chosen in the negative $x_{\mathrm{DF}}$ and $y_{\mathrm{DF}}$ semi--axes, respectively.}
\label{fig:photon_trajectory}
\end{figure*}

The crucial point for the derivation of the relationship between the complex visibility value and the number of counts $A$, $B$, $C$, and $D$ measured by the detector pixels, is the mathematical description of the sub--collimator transmission function. 
Since the transmission of the whole sub--collimator is the product of the transmission functions of the front and rear grids, we first consider a single window (either the front or the rear one).
The transmission function represents the probability that a photon, emitted from the point $(x,y)$ on the solar disk, hits the detector in the point $(w,h)$ on the corresponding detector surface. 
We note that the angular coordinates $x$ and $y$ actually represent the incident angles of the photon trajectory with respect to the $x_{\text{DF}}$ and $y_{\text{DF}}$ axes defined on the detector reference frame, respectively, as shown in Fig. \ref{fig:photon_trajectory}.
By using the approximation $\tan{z} \approx z$, which is valid in this context as the incident angles are small (usually less than $1$ degree), we have that the photon trajectory intersects the window $(x_{\text{WF}},y_{\text{WF}})$-plane in the point $(w + Lx, h + Ly)$, where $L$ denotes the distance between the window and the detector (see Fig. \ref{fig:photon_trajectory}).
Therefore, the transmission is equal to $1$ if $(w + Lx, h + Ly)$ falls inside a slit and $0$ otherwise.
The transmission of a single window is then a periodic function (with period equal to $p$) that is defined in terms of the distance between the point $(w + Lx, h + Ly)$ and the reference slit center (see the left panel of Fig. \ref{fig:photon_trajectory}) and is given by
\begin{equation}
\mathcal{T}(w,h,x,y) \coloneqq
\begin{cases}
1   & \text{if $-\frac{s}{2} +§ jp \leq \zeta - \varphi \leq \frac{s}{2} + jp$, $j\in\mathbb{Z}$}\\
0   & \text{otherwise}
\end{cases}
~,
\end{equation}
where
\begin{equation}\label{eq:w,h,x,y}
\zeta \coloneqq (w + Lx, h + Ly) \cdot \mathbf{k} ~.
\end{equation}
The function $\mathcal{T}$ is a bi--dimensional periodic square wave function.
Hence, we can approximate it by computing the corresponding Fourier series and by truncating the expansion at the fundamental harmonic.
Standard results of harmonic analysis \citep{gietal15,Siarkowski} lead to 
\begin{equation}\label{eq:transm_window}
\begin{split}
\mathcal{T}(w,h,x,y) &=\frac{s}{p} + \frac{2}{\pi} \sum_{n=1}^{+\infty}\frac{1}{n}\sin\left( \frac{n \pi s}{p} \right) \cos \left( 2\pi n \frac{\zeta - \varphi}{p} \right) \approx  \\
&\approx \frac{s}{p} + \frac{2}{\pi} \sin\left( \frac{\pi s}{p} \right) \cos \left( 2\pi \frac{\zeta  - \varphi}{p} \right) ~,
\end{split}
\end{equation}
where we approximated the transmission function by considering just the constant term and the fundamental harmonic.
As mentioned above, the transmission function of the whole sub--collimator, which will be denoted by $T$, can be defined as the product of the transmission functions $\mathcal{T}_f$ and $\mathcal{T}_r$ of the front and rear windows, respectively.
Specifically, $\mathcal{T}_f$ and $\mathcal{T}_r$ are obtained by replacing $s$, $p$, and $L$ in Eq. \eqref{eq:transm_window} with $s_f$, $p_f$, $d_{\text{sep}}+d_{\text{det}}$, and $s_r$, $p_r$, $d_{\text{det}}$, respectively.
Simple computations and approximations, which are described and discussed in Appendix \ref{appendix:sub-collimator_transm}, lead to the following expression for $T$:
\begin{equation}\label{eq: transmission}
\begin{split}
& T(w,h,x,y) \approx \frac{s_f s_r}{p_f p_r} + \\
&+\frac{2}{\pi^2} \sin \left( \frac{\pi s_f}{p_f} \right) \sin \left( \frac{\pi s_r}{p_r} \right) \cos \left(\frac{2\pi w}{W} + 2\pi (xu+yv) - c_{\varphi} \right) ~,
\end{split}
\end{equation}
\begin{figure*}[t]
\centering
\includegraphics[width=\textwidth]{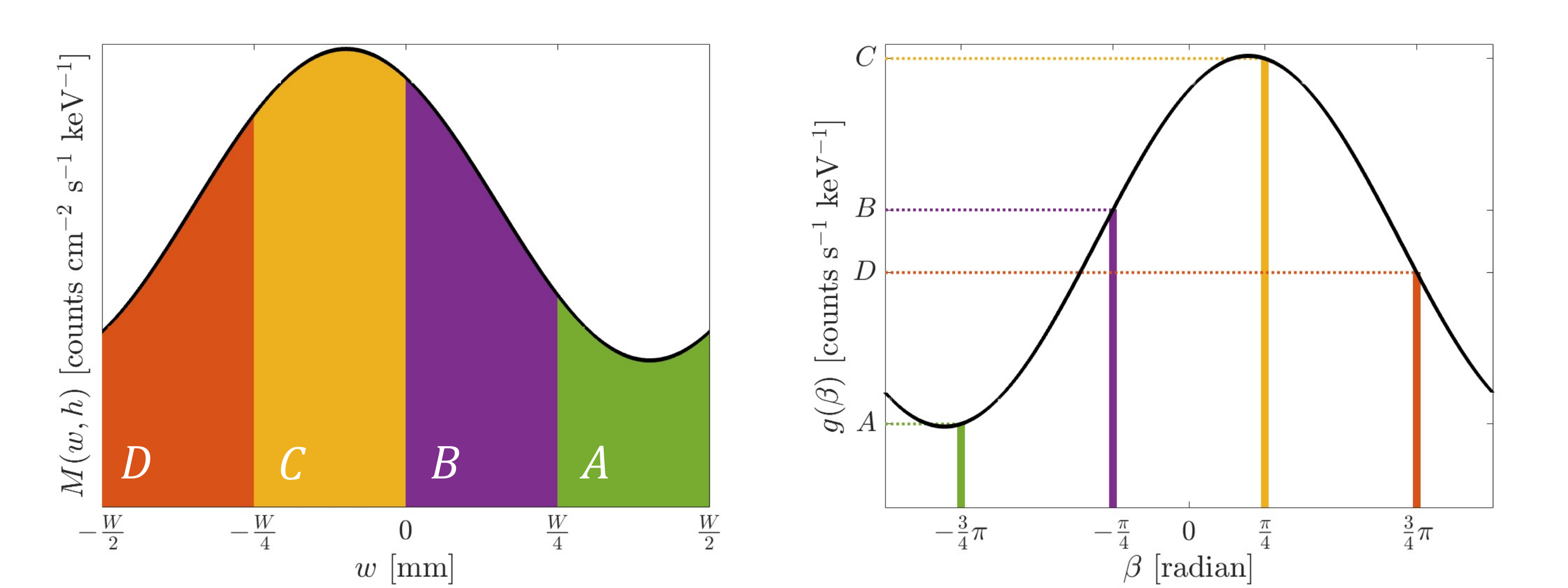}
\caption{\textit{Left panel:} profile of the (approximated) Moir\'e pattern shown in Fig. \ref{fig:moire-pattern} along the $w$-axis of the detector reference frame as seen looking towards the Sun. The colored areas represent the integrals of the pattern over the corresponding pixels (i.e., the number of counts recorded by each pixel). \textit{Right panel:} graph of $g(\beta)$ (Eq. \eqref{eq:g_beta}) over a period between $-\pi$ and $\pi$. The values of $g(\beta)$ in $ \beta = -\frac{3\pi}{4}, -\frac{\pi}{4}, \frac{\pi}{4}, \frac{3\pi}{4}$ are reported, showing that they are equal to $A$, $B$, $C$, and $D$, respectively.}
\label{fig:ABCD}
\end{figure*}
where
\begin{equation}\label{eq: def (u,v) and c}
(u,v) \coloneqq m\left[\frac{d_{\text{sep}}+d_{\text{det}}}{p_f} \mathbf{k}_f - \frac{d_{\text{det}}}{p_r} \mathbf{k}_r \right] ~,~ c_{\varphi} \coloneqq  2\pi m \left( \frac{\varphi_f}{p_f} - \frac{\varphi_r}{p_r} \right)~,
\end{equation}
and $m$ is the factor equal to $\pm 1$ at the right--hand--side of Eq. \eqref{eq: wave vector relations}.
Equation \eqref{eq: transmission} shows that the sub--collimator transmission function can be then approximated as the sum of a constant term, which represents the product of the mean transmission of the front and rear windows, and a sinusoidal wave with period equal to $W$ and orientation perpendicular to the detector vertical stripes.
The phase of the sinusoidal wave depends on the incident direction $(x,y)$ through its scalar product with the vector $(u,v)$, and on the relative phase between the front and the rear windows defined as $c_{\varphi}$.
Note that the phase of $T$ is then sensitive only to directions $(x,y)$ parallel to $(u,v)$, which, as we will show in the following, defines the angular frequency sampled by the sub--collimator.

Once derived the expression for the sub--collimator transmission function $T$ (Eq. \eqref{eq: transmission}), we can compute the transmitted X-ray photon flux as  
\begin{equation}\label{eq: def moire}
M(w,h) \coloneqq \iint \phi(x,y) T(w,h,x,y) \,dx\,dy ~.
\end{equation}
Equation \eqref{eq: def moire} represents the mathematical expression of the intensity of the Moir\'e pattern generated by the transmitted flux on the detector surface.
By replacing \eqref{eq: transmission} in \eqref{eq: def moire} and by computing the integral, one can prove that
\begin{equation}\label{eq: express moire}
\begin{split}
&M(w,h) \approx F \frac{s_f s_r}{p_f p_r} + \\
&+ \frac{2}{\pi^2} \sin \left( \frac{\pi s_f}{p_f} \right) \sin \left( \frac{\pi s_r}{p_r} \right) \mathcal{A} \cos\left(\frac{2\pi w}{W} + \omega - c_{\varphi} \right) ~,
\end{split}
\end{equation}
where 
\begin{equation}\label{eq:tot_flux}
F \coloneqq \iint \phi(x,y) \,dx \,dy
\end{equation}
is the total emitted flux and $\mathcal{A}$, $\omega$ are the amplitude and the phase of the visibility computed at the frequency $(u,v)$ defined in \eqref{eq: def (u,v) and c}.
Equation \eqref{eq: express moire} shows that the problem of determining amplitude and phase of the complex value of $V(u,v)$ is equivalent to determine amplitude and phase of the Moir\'e pattern from the number of counts measured by the detector pixels (see Fig. \ref{fig:moire-pattern}).

\begin{figure*}[ht]
\centering
\includegraphics[width=\textwidth]{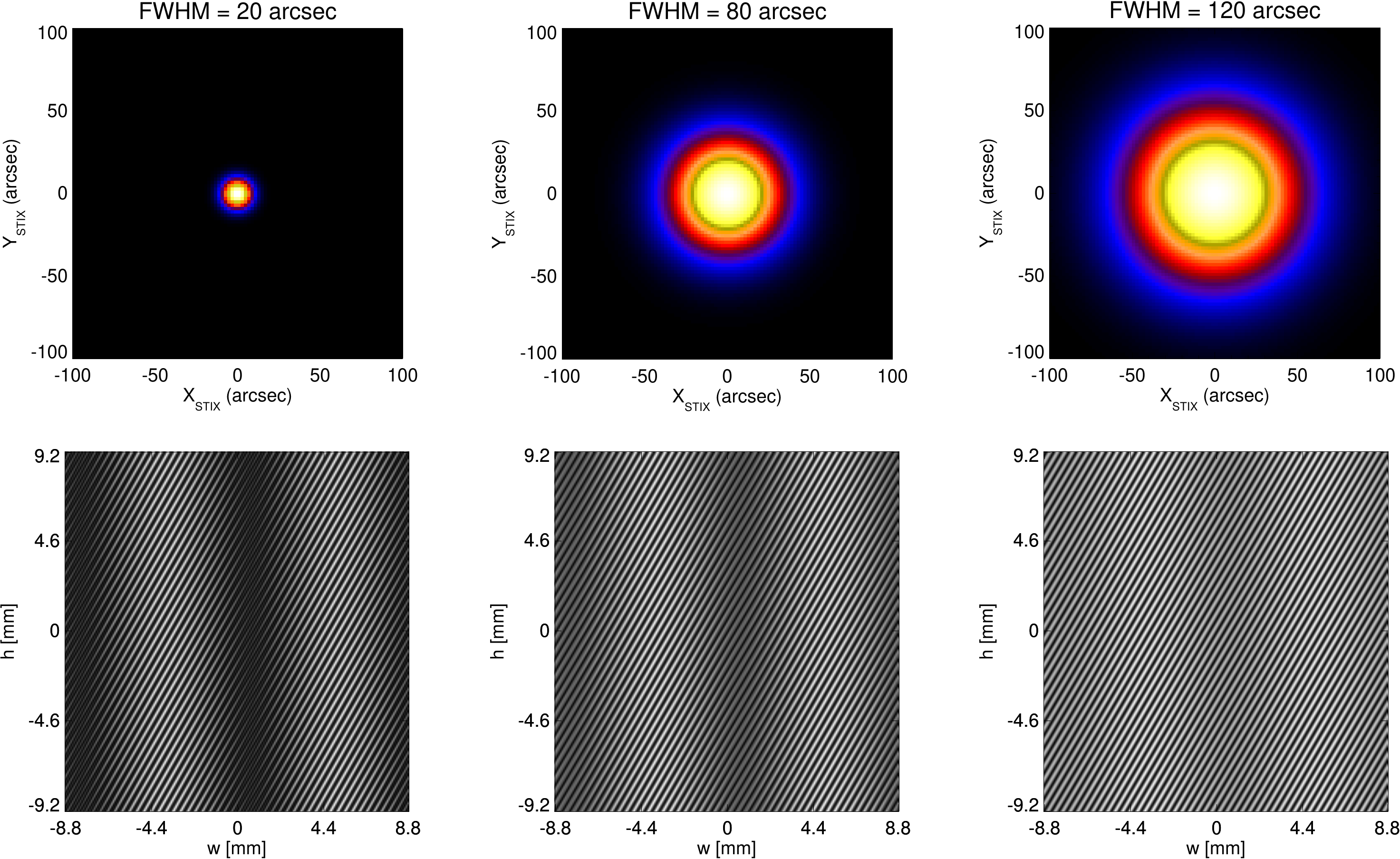}
\caption{\textit{Top row:} simulated circular Gaussian sources located in the center of the $(x,y)$-plane. The FWHM of the sources is equal to 20, 80 and 120 arcsec from left to right, respectively.
\textit{Bottom row:} simulated Moir\'e patterns created by sub--collimator 8 ($7b$, with a resolution of 61 arcsec) and corresponding to each one of the sources.
Note that the width of the plotted pattern is twice the width of the detector and that, consequently, two periods of the patterns are shown in the plots.}
\label{fig:moire_size}
\end{figure*}

The pixel counts $A$, $B$, $C$, and $D$ are obtained as the integral of the Moir\'e pattern $M$ over the corresponding pixel area. 
For the purpose of this description, we consider as pixels the four vertical stripes that constitute the detector and therefore we neglect the partition of each stripe into three units. 
Simple computations show that (see Appendix \ref{appendix:visib_amp_phase} for details)
\begin{equation}\label{eq: counts ABCD 1}
\begin{split}
A &= \mathcal{P} \left[F + \mathcal{E} \, \mathcal{A}\, \cos\left(\frac{3}{4}\pi + \omega - c_{\varphi}\right)\right] ~, \\
B &= \mathcal{P} \left[F + \mathcal{E} \, \mathcal{A}\, \cos\left(\frac{\pi}{4} + \omega - c_{\varphi} \right)\right] ~, \\
C &= \mathcal{P} \left[F + \mathcal{E} \, \mathcal{A}\, \cos\left(-\frac{\pi}{4} + \omega - c_{\varphi} \right)\right] ~, \\
D &= \mathcal{P} \left[F + \mathcal{E} \, \mathcal{A}\, \cos\left(-\frac{3}{4}\pi + \omega - c_{\varphi} \right)\right] ~.
\end{split}
\end{equation}
where
\begin{equation}\label{eq:p and e}
\mathcal{P}\coloneqq \frac{WH}{4} \frac{s_f s_r}{p_f p_r} ~,~ \mathcal{E} \coloneqq \frac{4\sqrt{2}}{\pi^3} \frac{p_f p_r}{s_f s_r} \sin\left( \frac{\pi s_f }{p_f} \right) \sin\left( \frac{\pi s_r }{p_r} \right)
\end{equation}
are the pixel area multiplied by the mean transmission of the sub--collimator, and the \textit{modulation efficiency} of the fundamental harmonic, respectively.
The modulation efficiency describes what percentage of the flux of an ideal point--like source is modulated by the fundamental harmonic of the transmission function of the windows; in the case of the STIX sub--collimators, $\mathcal{E}\approx 0.7$.
It is worth noticing that the pixel counts $A$, $B$, $C$, and $D$ represent the values of the function 
\begin{equation}\label{eq:g_beta}
g(\beta) =  \mathcal{P} \left[F + \mathcal{E}\, \mathcal{A}\, \cos\left(-\beta + \omega - c_{\varphi} \right)\right]   
\end{equation}
computed in $\beta = -\frac{3\pi}{4}, -\frac{\pi}{4}, \frac{\pi}{4}, \frac{3\pi}{4}$, respectively, as shown in Fig. \ref{fig:ABCD}.
Therefore, in principle, one could retrieve the values of the visibility amplitude and phase by forward--fitting $F$, $\mathcal{A}$, and $\omega$ from $A$, $B$, $C$, and $D$ using Eq. \eqref{eq:g_beta}.
However, as described in Appendix \ref{appendix:visib_amp_phase}, it is possible to derive an analytical expression for the relationship between the pixel counts and the visibility amplitude and phase, which is given by
\begin{equation}\label{eq:vis_amp_phase}
\begin{split}
\mathcal{A} &= \frac{1}{2\,\mathcal{P}\,\mathcal{E}} \sqrt{ (C-A)^2 + (D-B)^2 } \\
\omega      &= \mathrm{arctan2} \left(D-B,C-A\right) + \frac{\pi}{4} + c_{\varphi}
\end{split}
~,
\end{equation}
where $\mathrm{arctan2}(b,a)$ associates to $(a,b)$ the corresponding angle in polar coordinates.
Consequently, the complex visibility value sampled by the considered sub--collimator is
\begin{equation}\label{eq:vis}
V(u,v) = \mathcal{A} \left(\cos(\omega) + i \sin(\omega) \right) ~,
\end{equation}
where $(u,v)$ and $\mathcal{A}$, $\omega$ are defined as in Eqs.  \eqref{eq: def (u,v) and c} and \eqref{eq:vis_amp_phase}, respectively.

We conclude this section by providing a visual explanation of how the amplitude and phase of a Moir\'e pattern are related to the amplitude and phase of the corresponding visibility.
We recall that a visibility amplitude is related to the morphology of the X--ray source.
Specifically, it contains information on the source dimension along a direction \textit{parallel} to the corresponding $(u,v)$ frequency. 
On the other hand, the value of the visibility phase is determined by the projection of the location of the source along the same direction.

In the top--row panels of Fig. \ref{fig:moire_size} we show three simulated circular Gaussian sources centered at the origin of the $(x_{\text{STIX}},y_{\text{STIX}})$ coordinate system on the solar disk. 
The Full Width at Half Maximum (FWHM) of the sources increases from left to right, taking values equal to $20$, $80,$ and $120$ arcsec, respectively.
In the bottom--row panels of the same figure we show the simulated Moir\'e patterns created by sub--collimator $8$ ($7b$) and corresponding to each one of the sources. 
The pattern amplitude clearly decreases with the increasing size of the source; in particular, the pattern is almost flat, its amplitude being close to zero, when the source size is larger than the resolution\footnote{The sub--collimator resolution is defined as $1/(2\Vert (u,v)\Vert)$, where $(u,v)$ is the sampling frequency and $\Vert \cdot \Vert$ denotes the Euclidean norm.} of the sub--collimator, which is 61 arcsec (bottom--right panel).
This is consistent with the fact that the amplitude of the Fourier transform of a Gaussian function is another Gaussian function defined in the $(u,v)$-plane, whose FWHM is inversely proportional to the one of the original function defined in the $(x,y)$-plane. 
Hence, when the FWHM of the X--ray source increases, the corresponding Fourier amplitude FWHM becomes narrower and, consequently, the value of the amplitude of the visibility sampled by the sub--collimator decreases to zero.

\begin{figure*}[ht]
\centering
\includegraphics[width=\textwidth]{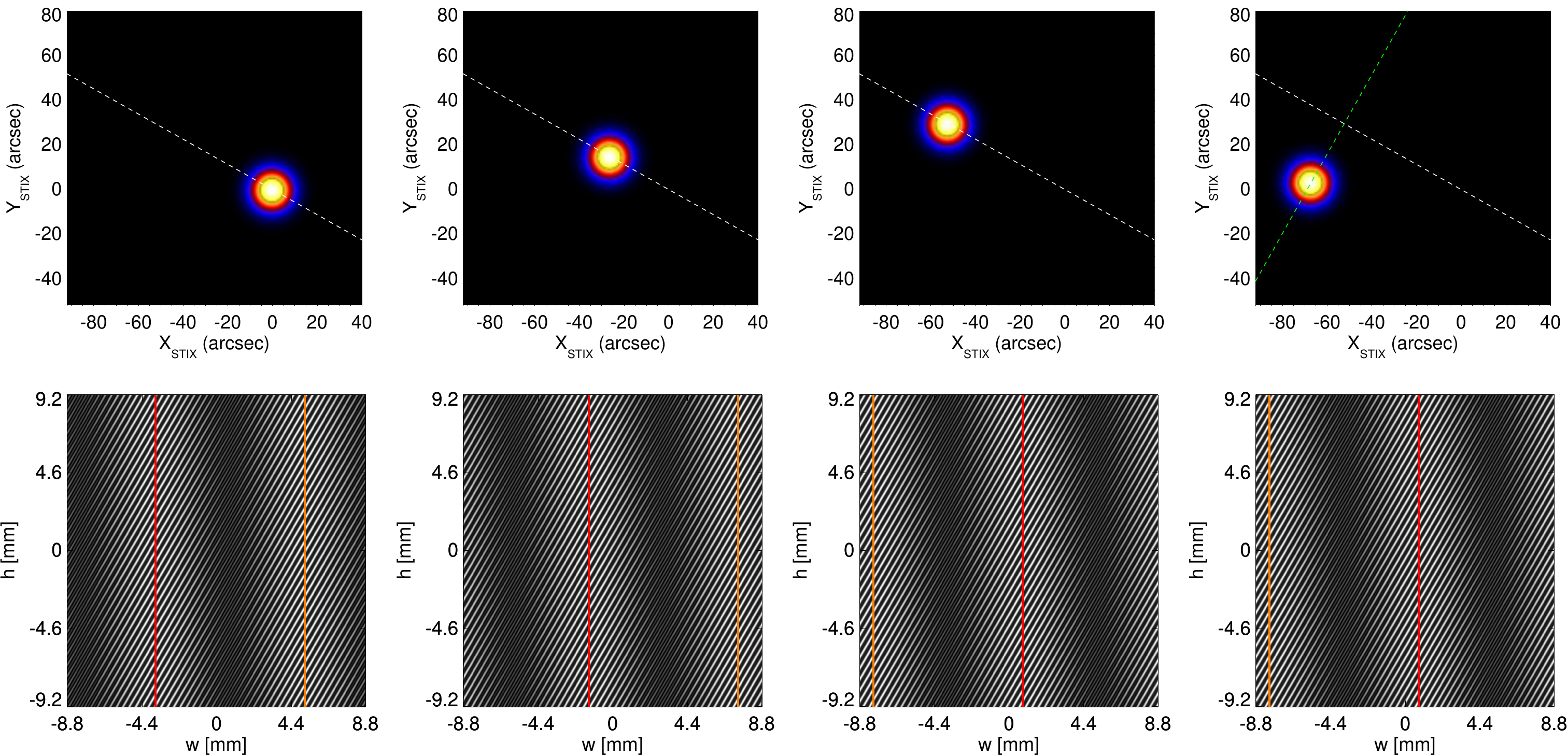}
\caption{\textit{Top row:} simulated circular Gaussian source with FWHM equal to 20 arcsec. From left to right, the source location is shifted from $(x,y)=(0,0)$ by 30 and 60 arcsec along the direction determined by the $(u,v)$ frequency sampled by sub-collimator 8 ($7b$). 
Such a direction is indicated with a white dashed line. 
In the top right panel the source is further shifted by 30 arcsec along a direction orthogonal to $(u,v)$, which is plotted with a green dashed line.
\textit{Bottom row:} Moir\'e patterns created by sub--collimator 8 ($7b$) and corresponding to each one of the circular Gaussian sources.
The red and the orange vertical lines denote two consecutive peaks of the pattern. 
Note that the width of the plotted patterns is twice the width of the detector and that, consequently, two periods of the patterns are shown in the plots.}
\label{fig:moire_location}
\end{figure*}

Similarly, in the top--panels of Fig. \ref{fig:moire_location} we show a simulated circular Gaussian source with FWHM equal to 20 arcsec and located in different positions within the FOV. 
Starting from $(x,y)=(0,0)$ (top--left panel), the source is shifted by 30 and 60 arcsec along a direction parallel to the $(u,v)$ frequency sampled by sub--collimator $8$ (second and third top--row panels from the left, respectively).
In the top--right panel of the same figure, the source location is shifted by 30 arcsec with respect to the previous position, but this time along a direction perpendicular to $(u,v)$.
In the bottom panels we show the simulated Moir\'e patterns generated by sub--collimator 8 ($7b$) and corresponding to the source in the different locations.
The width of the patterns that are shown in the bottom panels is twice as the width of the detector; therefore, two periods of the patterns are reported in the panels.
We highlight with a red and an orange line two consecutive peaks of the pattern, whose distance from the center of the detector is related to the pattern phase.
We notice that the location of the peaks of the Moir\'e pattern (and, hence, its phase) changes when the source moves along a direction parallel to $(u,v)$ (first three bottom--row panels from the left).
In particular, due to the periodicity of the pattern, the right peak in the second panel from the left, which is highlighted in orange, appears close to the left edge in the third panel.
On the contrary, the pattern phase in the third and fourth bottom--row panels is the same, since the source moves perpendicularly to $(u,v)$.
We can then appreciate how the pattern phase, which is closely related to the visibility phase, depends only on the projection of the source location along the direction determined by $(u,v)$. 

\subsection{Count distribution model}

We now describe the mathematical model that links the photon flux $\phi$ directly to the number of counts $A$, $B$, $C$, and $D$ recorded by the detector pixel \citep{massa2019count,Siarkowski}.
Essentially, we derive the expression of the \textit{modulation pattern} of the sub-collimator pixels \citep{2002SoPh..210...61H}, which represents the probability that a photon emitted from a point $(x,y)$ on the solar disk is recorded by a specific detector pixel.

We denote by $N$ the number of counts recorded by a detector pixel ($N$ can be either $A$, $B$, $C$, or $D$). 
As we discussed in the previous section, $N$ is given by the integral of the pattern over the corresponding pixel area $P$. 
Starting from Eq. \eqref{eq: def moire}, we obtain
\begin{equation}\label{eq:n_counts}
\begin{split}
N &= \iint_{P} M(w,h) \,dw \,dh  =\\
&=\iint_{P} \left[ \iint \phi(x,y) T(w,h,x,y) \,dx\,dy \right] \,dw \,dh ~.
\end{split}
\end{equation}
By inverting the order of integration in \eqref{eq:n_counts}, by using the analytical expression of the transmission defined in Eq. \eqref{eq: transmission} and by using Eq. \eqref{eq:magic_formula} derived in Appendix \ref{appendix:visib_amp_phase} (in which we replace $\omega$ with $2\pi(xu+yv)$), one can prove that
\begin{equation}\label{eq:count_formation_model}
\begin{split}
A &= \iint \phi(x,y) \left[\mathcal{P}\left(1 - \mathcal{E} \cos(q(x,y)) \right) \right] \,dx \,dy, \\
B &= \iint \phi(x,y) \left[\mathcal{P}\left(1 - \mathcal{E} \sin(q(x,y)) \right)\right] \,dx \,dy, \\
C &= \iint \phi(x,y) \left[\mathcal{P}\left(1 +\mathcal{E} \cos(q(x,y))\right) \right] \,dx \,dy, \\
D &= \iint \phi(x,y) \left[\mathcal{P}\left(1 + \mathcal{E} \sin(q(x,y))\right) \right] \,dx \,dy ~,
\end{split}
\end{equation}
where
\begin{equation}\label{eq:count_q}
q(x,y) \coloneqq 2\pi \left(xu+yv\right) -\frac{\pi}{4} - c_{\varphi} ~,
\end{equation}
and $(u,v)$, $c_{\varphi}$ and $\mathcal{P}$, $\mathcal{E}$ are defined as in Eqs. \eqref{eq: def (u,v) and c} and \eqref{eq:p and e}, respectively.
The pixel modulation patterns are then the functions inside the integrals in \eqref{eq:count_formation_model} that multiply the photon flux $\phi$.
They consist of the sum of a constant term (the average transmission) and a sinusoidal function whose period and orientation are determined by the $(u,v)$ point sampled by the sub--collimator.
The integral equations in \eqref{eq:count_formation_model} represent the mathematical model that links the photon flux to the pixel counts and allows addressing the STIX imaging problem directly from the counts themselves, without computing the complex visibility values.

\subsection{Discussion}\label{section:discussion}

We conclude this section with a few comments.

\begin{itemize}

\item In Sect. \ref{section:visibility} we proved that each one of the STIX sub--collimators measures a visibility corresponding to the $(u,v)$ frequency defined in \eqref{eq: def (u,v) and c}.
By design, the frequencies are located on 10 concentric circles in the $(u,v)$-plane; each circle contains three frequencies, which are sampled by sub--collimators labelled with the same number and a different letter.
The design of the sub--collimators is done so that the center of the circles coincides with the origin of the $(u,v)$--plane if the detectors are right behind the rear grid.
As the nominal separation between the rear grid and the detectors is 4.7 mm, the center of the circles results to be slightly shifted in the $u$ direction, its coordinates being $($\(\approx \)$1.8 \times 10^{-4} \,\mathrm{arcsec}^{-1},0\, \mathrm{arcsec}^{-1})$, and the circle radii are only approximately inversely proportional to the sub--collimator resolutions.
The instrument measures 10 different resolutions, from \(\approx \)$7$ to \(\approx \)$180$ arcsec, logarithmically spaced in steps of \(\approx \)$1.43$.
The STIX $(u,v)$ coverage is shown in Fig. \ref{fig:uv_points}, while the values of the resolution of each sub--collimator and of the orientation angle of the corresponding visibilities are reported in Table \ref{tab:res_orient}.
\begin{figure}[ht]
\centering
\includegraphics[width=0.8\columnwidth, angle=90]{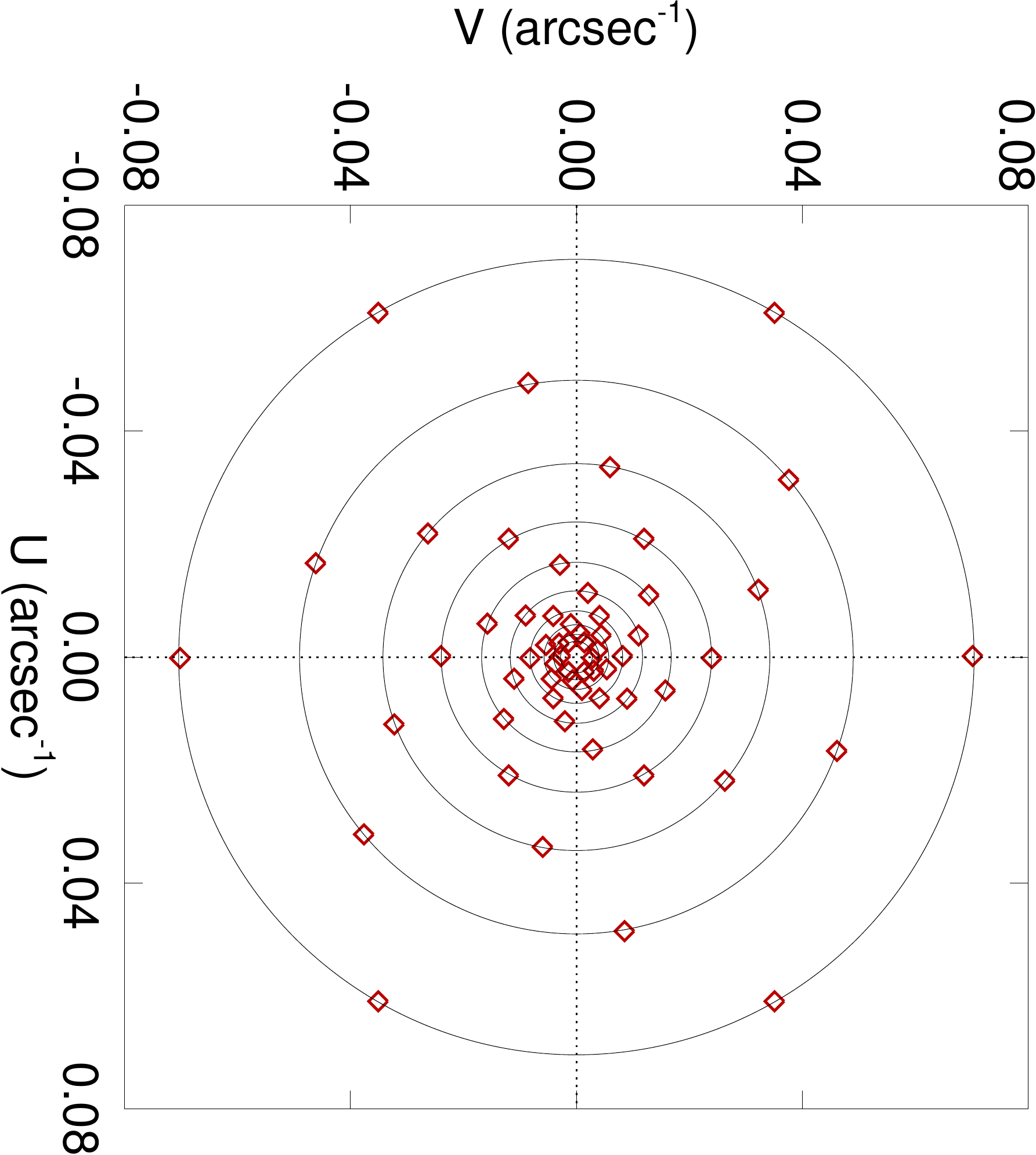}
\caption{Frequencies sampled by the STIX sub--collimators (red dots). 
As a reference, the ten circles on which the frequencies lie are plotted in black.
Note that the STIX $(u,v)$--coverage has been mirrored by considering the symmetrical of the frequencies with respect to the origin of the $(u,v)$--plane.
Hence, 60 frequencies are plotted in order to better show the characteristic spiral configuration of the STIX $(u,v)$--coverage.}
\label{fig:uv_points}
\end{figure}
Note that, since the function $\phi$ representing the intensity of the photon flux is a real valued function, the following relation holds true for each complex visibility value measured by STIX:
\begin{equation}\label{eq:conj_V}
V(-u,-v) = \overline{V(u,v)} ~,
\end{equation}
where the bar denotes the complex conjugate. Therefore, for each visibility value measured by STIX in correspondence to the $(u,v)$ point defined in \eqref{eq: def (u,v) and c}, we can retrieve the value of the visibility corresponding to $(-u,-v)$ by using \eqref{eq:conj_V}. 
It is then possible to extend the STIX $(u,v)$ coverage to 60 angular frequencies (as shown in Fig. \ref{fig:uv_points}) by considering the ones that are symmetric with respect to the origin of the $(u,v)$--plane.
However, symmetrically replicated visibilities contain redundant information, and including them in the image reconstruction process does not improve the quality of the reconstructed image.

\begin{table*}[t]
\centering
\begin{tabular}{ccccccc}
\toprule
&\multicolumn{2}{c}{$a$}    &\multicolumn{2}{c}{$b$}    &\multicolumn{2}{c}{$c$}\\
\cmidrule(l){2-3}
\cmidrule(l){4-5}
\cmidrule(l){6-7}
&\Longunderstack{\textbf{Resolution} \\ \textbf{[arcsec]}}   &\Longunderstack{\textbf{Orientation} \\ \textbf{[deg]}}  &\Longunderstack{\textbf{Resolution} \\ \textbf{[arcsec]}}   &\Longunderstack{\textbf{Orientation} \\ \textbf{[deg]}}  &\Longunderstack{\textbf{Resolution} \\ \textbf{[arcsec]}}   &\Longunderstack{\textbf{Orientation} \\ \textbf{[deg]}}\\
\midrule

10   &169.3   &-28.3   &178.3   &86.4   &169.3   &28.3\\
9   &119.6   &-9.6   &126.7   &107.6   &128.4   &-128.0\\
8   &84.7   &9.7   &85.6   &-48.7   &86.3   &68.4\\
7   &59.9   &29.4   &59.9   &-29.4   &61.0   &-88.8\\
6   &43.1   &-129.3   &43.3   &169.8   &42.9   &109.2\\
5   &29.7   &69.4   &30.1   &-169.9   &29.6   &-49.5\\
4   &20.9   &89.6   &20.7   &29.8   &20.7   &-29.8\\
3   &14.5   &-69.7   &14.6   &-129.8   &14.6   &169.9\\
2   &10.2   &129.8   &10.2   &-109.8   &10.1   &10.0\\
1   &7.1   &-29.9   &7.1   &-89.9   &7.1   &29.9\\
\bottomrule
\end{tabular}
\caption{Resolution of the STIX sub--collimators and orientation angle of the corresponding $(u,v)$ frequencies.
The number of each sub--collimator label is reported in the first column, while the letter of the label is reported in the first row.
We note that the orientation angle is computed counterclockwise from the positive $u$ semi-axis.}
\label{tab:res_orient}
\end{table*}

\item The instrument Point Spread Function (PSF) can be determined by setting each visibility value equal to 1 and by performing a discrete inverse Fourier transform. 
We can use different sets of weights for the discretization of the Fourier transform. 
With \textit{natural weighting} we give each visibility a weight value equal to 1; with \textit{uniform weighting}, instead, we give each visibility a weight proportional to the area of the $(u,v)$-plane that is represented by the corresponding $(u,v)$ frequency \citep{pianabook}.
Specifically, as it is shown in Fig. \ref{fig:uv_points}, the STIX $(u,v)$ coverage becomes more sparse in the part of the $(u,v)$-plane corresponding to high frequencies ($\Vert (u,v) \Vert \gtrsim 0.04$ arcsec$^{-1}$). Hence, with uniform weighting we give each visibility a weight equal to the distance of the $(u,v)$ frequency from the origin of the $(u,v)$-plane.
The left and the right panel of Fig. \ref{fig:PSFs} show the PSFs obtained with natural weighting and uniform weighting, respectively.
We can appreciate how the central beam of the PSF becomes narrower when uniform weighting is applied, the corresponding FWHM being equal to \(\sim \)$11$ arcsec instead of \(\sim \)$19$ arcsec as in the case of natural weighting.
Hence, applying uniform weighting increases the resolution of the reconstructed image. 
On the other hand, due to the relatively larger sidelobes that are present in the PSF obtained with uniform weighting, the images usually have a lower dynamic range.

\begin{figure*}[ht]
\centering
\includegraphics[width=0.8\textwidth]{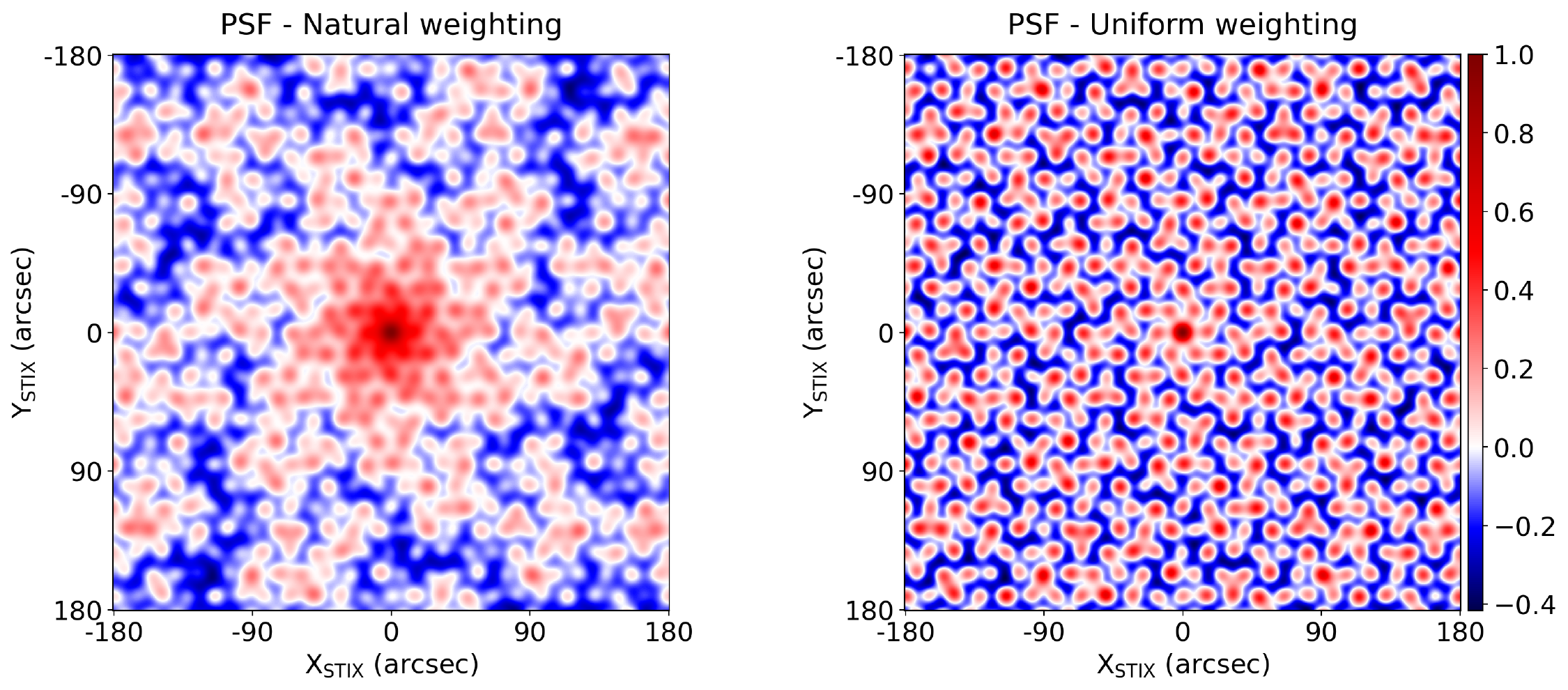}
\caption{STIX Point Spread Function (PSF) obtained by applying natural and uniform weighting (left and right panel, respectively).
All 30 frequencies sampled by the imaging sub--collimators are considered for computing the PSFs.
The same color map is shared by the two plots of the PSF.
Note that the 0 value corresponds to ``white.''
}
\label{fig:PSFs}
\end{figure*}

\item As stated at the beginning of Sect. \ref{section:data-formation}, for deriving Eq. \eqref{eq:vis_amp_phase}, which describes the link between the pixel counts and the visibility amplitude and phase, we considered as pixels each one of the four vertical stripes making a STIX detector.
However, it is possible to compute the visibility values using only top--row pixels, bottom--row pixels or small pixels,
as long as slight modifications to the derived formulas are made.
In particular: 1) the factor $\mathcal{P}$ containing the value of the pixel area has to be adapted based on the selected pixels; and 2) the 45 degree angle ($\pi/4$) in the visibility phase formula becomes 46.1 degrees if only top--row pixels or bottom--row pixels are selected, and 22.5 degrees if only small pixels are selected.
Specifically, the value of 46.1 degrees is obtained by taking into account that the shape of top--row and bottom--row pixels is not exactly rectangular (due to the presence of the small central pixel; see the right panel of Fig. \ref{fig:scheme-grid}). Instead, the value of 22.5 degrees (which is the half of 45 degrees) is a consequence of the fact that the width of a small pixel is half the width of the detector vertical stripe.
We note that analogous modifications have to be performed to the count distribution model (Eqs. \eqref{eq:count_formation_model} and \eqref{eq:count_q}) depending on the selected configuration of pixels.

\item The STIX imaging technique would work also in the case the detectors were partitioned in a number of identical vertical stripes different from four, provided that appropriate modifications are performed to Eqs. \eqref{eq: counts ABCD 1} through \eqref{eq:g_beta} and to Eqs. \eqref{eq:count_formation_model} and \eqref{eq:count_q}.
As for the visibility formation model, since the values of $F$, $\mathcal{A}$, and $\omega$ are to be determined from the experimental pixel counts, at least three measurements of the Moir\'e pattern (and, hence, three pixels) are needed.
The choice of four pixels adopted for STIX is a compromise between data redundancy and complexity of the detector design.
In the case the number of pixel stripes was different from four and an analytical expression of the link between the number of counts and the visibility amplitude and phase (Eq. \eqref{eq:vis_amp_phase}) would be difficult to be determined, it would be always possible to retrieve the values of $\mathcal{A}$ and $\omega$ by applying a forward--fitting method to a modified version of the function $g(\beta)$ defined in \eqref{eq:g_beta}.

\end{itemize}

Finally, we briefly compare strengths and limitations of the visibility formation process and of the count distribution model.
Visibilities are not sensitive to a constant background radiation included in the measured data, since both amplitudes and phases are defined in terms of subtractions between counts.
Assuming that the detected background radiation is the same in $A$, $B$, $C$, and $D$, then it cancels out thanks to subtraction.
This is not the same for the pixel counts: for addressing the image reconstruction problem directly from $A$, $B$, $C$, and $D$, an accurate background subtraction needs to be performed before applying any reconstruction method.
This is particularly crucial for those events in which the intensity of the background is comparable to that of the signal originated from the flaring source.
Moreover, visibilities allow the use of Fast Fourier Transform (FFT) during the image reconstruction process, thus increasing the speed of the image  reconstruction algorithms.
On the other hand, image reconstruction from pixel counts takes advantage of the fact that data are more numerous (120 pixel counts vs 60 real and imaginary parts of visibilities), although the information content of counts and visibilities is the same due to redundancy. Also, count measurements are mainly affected by Poisson noise, while real and imaginary parts of visibilities follow a Skellam statistics because of the differences between counts. 
Hence, the S/N of counts is higher than the one of the real and imaginary parts of the visibilities \citep{massa2019count}.

\section{STIX Fields-of-View}\label{section:def_fov}

\begin{figure*}[ht]
\centering
\includegraphics[width=0.8\textwidth]{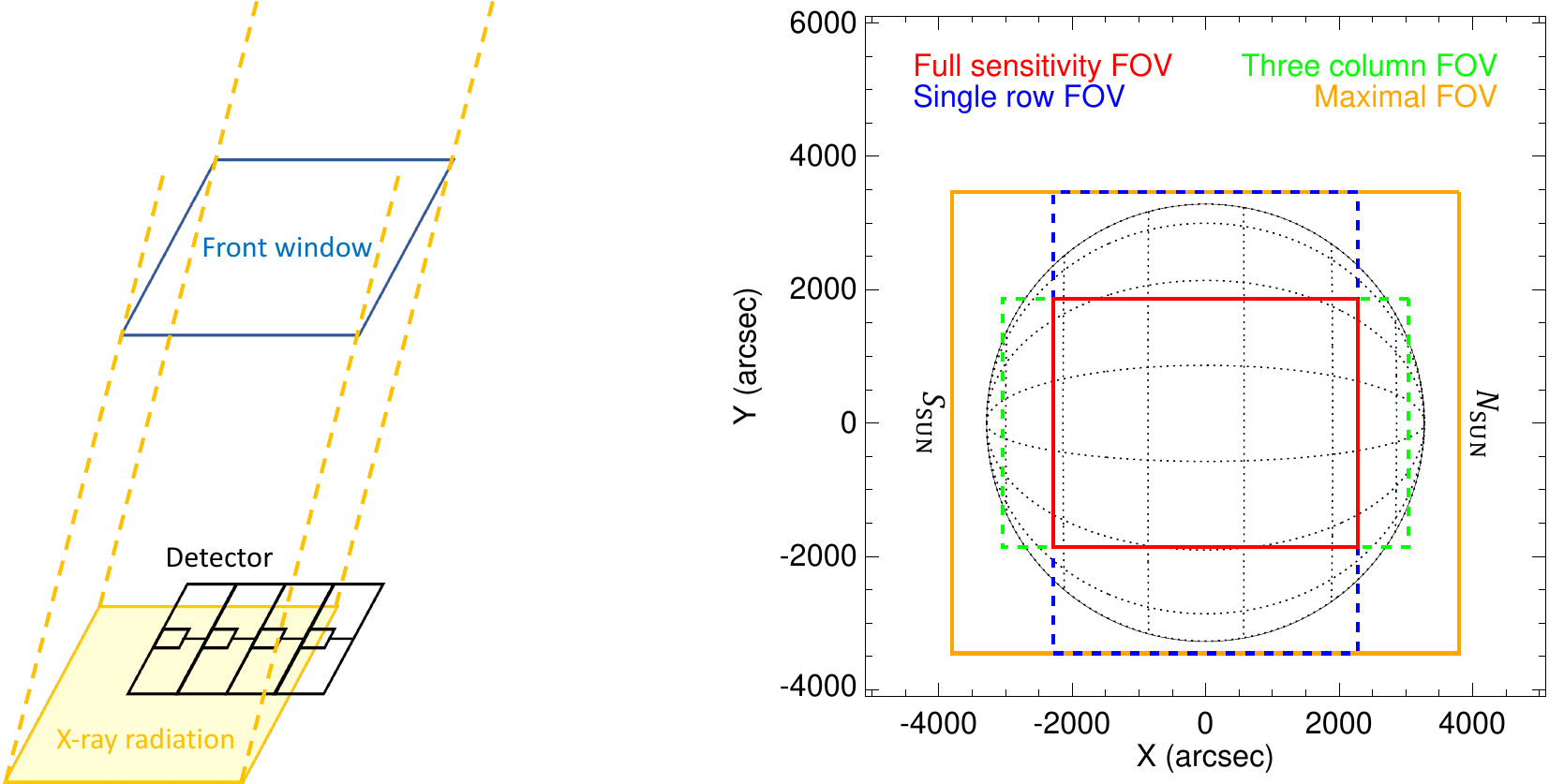}
\caption{\textit{Left panel:} schematic of the rectangular area of the front window that is projected by the flaring X--ray radiation on the detector plane, in the case of an off--axis source. Note that the detector is not fully illuminated by the X--ray radiation. \textit{Right panel:} different STIX FOVs overlaid to a solar disk with apparent radius of 3280 arcsec as it appeared during the Solar Orbiter perihelion of October 12, 2022, when the distance between the spacecraft and the Sun was \(\sim \)$0.29$ AU.
The full--sensitivity FOV, single row FOV, three column FOV, and maximal FOV are plotted in red, blue, green, and orange, respectively.
Note that the Sun is plotted 90 deg rotated as observed by STIX (see Eq. \eqref{eq:eq_rot_stix_sun} and Fig. \ref{fig:reference_frames}).}
\label{fig:FOVs_1}
\end{figure*}

The STIX FOV for imaging\footnote{In this section we discuss the definition of the instrument FOV, and not the FOV adopted for image reconstruction (i.e., the size of the image that is reconstructed from STIX data), which can be arbitrarily selected by the user.} is constrained solely by the separation between the front grid and the detectors (nominally, 597 mm), and by size of the front grid windows (nominally, 22 mm $\times$ 20 mm).
For off--axis sources, specific detector pixels, which are the same in each sub--collimator, are not fully illuminated by the X--ray radiation and have to be discarded for the computation of the visibilities or for direct image reconstruction using Eq. \eqref{eq:count_formation_model}.
This happens when the radiation passing through the front window of the sub--collimator creates an illuminated rectangular area (the projection of the front window on the detector plane) that does not completely contain the detector, as shown in left panel of Fig. \ref{fig:FOVs_1}.

Based on the configuration of pixels that are fully illumimated by the X--ray radiation, we can define different FOVs (we refer the reader to Appendix \ref{appendix:fovs_definition} for more details). 
The \textit{full sensitivity} FOV, with all pixels fully illuminated, has dimension $1.267$ deg $\times$ $1.036$ deg, and it is plotted in red in the right panel of Fig. \ref{fig:FOVs_1}.
This FOV does not contain the entire solar disk when the instrument is close to the Sun (as in the case of the Solar Orbiter perihelion of October 12, 2022, when the distance between the spacecraft and the star was \(\sim \)$0.29$ AU and the apparent solar radius was \(\sim \)$3280$ arcsec; see the right panel of Fig. \ref{fig:FOVs_1}), or when the spacecraft is pointing far off the Sun center, for example for limb--pointing.
Although some effective area is lost, imaging is still possible if only some pixels are fully illuminated.  
Specifically, for source offsets in the $y_{\text{STIX}}$ direction, imaging with full fidelity requires only one fully-illuminated row of pixels, and this allows expanding the FOV to 1.919 deg in the $y_{\text{STIX}}$ direction.
The resulting \textit{single row} FOV is reported with a blue dashed line in the right panel of Fig. \ref{fig:FOVs_1}.   
For source offsets in the $x_{\text{STIX}}$ direction, if one of the outer columns of pixels is not fully illuminated,  image reconstruction can still be performed using the counts recorded by the other three\footnote{From Eq. \eqref{eq: counts ABCD 2} in Appendix \ref{appendix:visib_amp_phase} it is straighforward to prove that, if $A$ is discarded, then amplitude and phase of the visibilities can be obtained by replacing $C-A$ with $2C - (B+D)$ in \eqref{eq:vis_amp_phase}. 
On the contrary, if $D$ is discarded, then $D-B$ has to be replaced with $A+C-2B$ in the same equation.},  albeit with some sacrifice in redundancy, and effective area. In this case, the FOV expands to 1.689 deg in the $x_{\text{STIX}}$ direction, and it is named \textit{three column} FOV (green dashed line in the right panel of Fig. \ref{fig:FOVs_1}).
The STIX FOV can be further extended in the same direction up to 2.111 deg if two adjacent pixels columns are fully illuminated. 
Imaging can make use of the pixel pair to provide a non-redundant sine-cosine response similar to that used by with HXT and HXI, although, in this case, an estimate of the total flux of the source is needed for correctly computing the visibility values.
The \textit{maximal} FOV, which is defined considering a single row and two columns of pixels, is reported in orange in the right panel of Fig \ref{fig:FOVs_1}.
Note that this FOV encompasses the full solar diameter at perihelion.   

In determining the FOVs, it should be noted that each rear grid window (nominally, 13 mm $\times$ 13 mm) is slightly oversized so as to not represent a constraint in itself.   It should also be noted that for large source offsets, the relatively narrow (3 mm) separation between windows can enable some photons to go through the front window of one subcollimator and the rear window of another.
By considering only fully-illuminated pixels, however, this does not affect imaging. 
With careful analysis, this effect can extend the FOV for spatially-integrated spectroscopy beyond that for imaging.
For determining which rows or columns of pixels have to be discarded for imaging a specific flaring source, we can exploit either information on the source location provided by the CFL \citep{krucker2020spectrometer} or information derived from data redundancy, as it will be discussed in the third paper of the series.

From a practical viewpoint, almost all STIX flares are detected within the full sensitivity FOV. A few flares observed near perihelion occurred within the single row imaging FOV (e.g. SOL2022-03-30T18), resulting in a loss of sensitivity of about half the counts. To date, no flare has been detected outside the single row imaging FOV, but it might happen during one of Solar Orbiter's special observing runs when the spacecraft is pointing at the solar polar regions near perihelion, and a flare occurs at high latitude on the other hemisphere.

\section{Image reconstruction problem and imaging methods}\label{section:imging_methods}

The image reconstruction problem for STIX can be formulated in two different ways, depending on the kind of data that are considered (either visibilities or counts). 
The imaging problem from visibilities consists in determining the image of the flaring source $\phi$ which satisfies the following equation
\begin{equation}\label{eq:imaging_problem_vis}
\mathcal{F}\phi = \mathbf{V}~,
\end{equation}
where $\mathcal{F}$ is the Fourier transform computed at the $(u,v)$ frequencies sampled by STIX (see Eq. \eqref{eq: def (u,v) and c} and Fig. \ref{fig:uv_points}) and $\mathbf{V}$ is the array containing the experimental visibility values defined in \eqref{eq:vis}.
Differently, the image reconstruction problem can be addressed directly from pixel counts by solving 
\begin{equation}\label{eq:imaging_problem_counts}
\mathcal{H} \phi = \mathbf{C}~,
\end{equation}
where $\mathcal{H}$ comes from the discretization of the integrals defined in \eqref{eq:count_formation_model}, and $\mathbf{C}$ is the array whose entries are the counts recorded by the pixels in each detector.
In both cases, the STIX imaging problem is an ill--posed inverse problem which has to be addressed by means of regularization techniques \citep{pianabook} for taking into account the non--uniqueness of the solution and for mitigating the effects of noise amplification during the reconstruction process.

Below, we provide an overview of the imaging methods that are currently implemented in the Interactive Data Language (IDL) STIX ground software \citep{ewan_dickson_2022_7277180} for the solution of \eqref{eq:imaging_problem_vis} and \eqref{eq:imaging_problem_counts}.
For a comprehensive review of the different methods and their parameters, we refer the reader to \cite{massa2022hard} and \cite{pianabook}.

\subsection{Back Projection}
This is the simplest (and most naive) visibility-based method, given that it implements a direct discrete Fourier transform inversion of STIX visibilities \citep[e.g.,][]{2002SoPh..210...61H}. 
Similarly to what discussed in Sect. \ref{section:discussion}, the Fourier integration can be computed using different quadrature formulae (i.e., different weights in the numerical integration scheme).
In particular, we shall consider below both natural and uniform weighting.

\subsection{CLEAN}
The main ingredients for this iterative deconvolution algorithm \citep{1974A&AS...15..417H,2009ApJ...698.2131D} are the back-projected image and the dirty beam (i.e., the instrument PSF). The CLEAN loop iteratively identifies a point source in the back-projected image, saves its intensity and position in a CLEAN components map and updates the back-projected map by subtracting the dirty beam positioned at the identified point source.
The reconstructed map is obtained by convolving the final CLEAN components map with an idealized version of the PSF. 

\subsection{Expectation Maximization (EM)}
At the moment this is the only method implemented in the STIX ground software that is fed with STIX count rates \citep{massa2019count,benvenuto2013expectation}, although another EM-like approach \citep{Siarkowski} will be released in the future. EM is an iterative scheme to realize likelihood maximization under the constraints that the pixel content must be non--negative and the noise affecting the input data must follow the Poisson statistics.
The stopping rule for this implementation of EM is described in \cite{2014InvPr..30c5012B}. 

\begin{figure*}[!p]
\centering
\includegraphics[height=0.92\textheight]{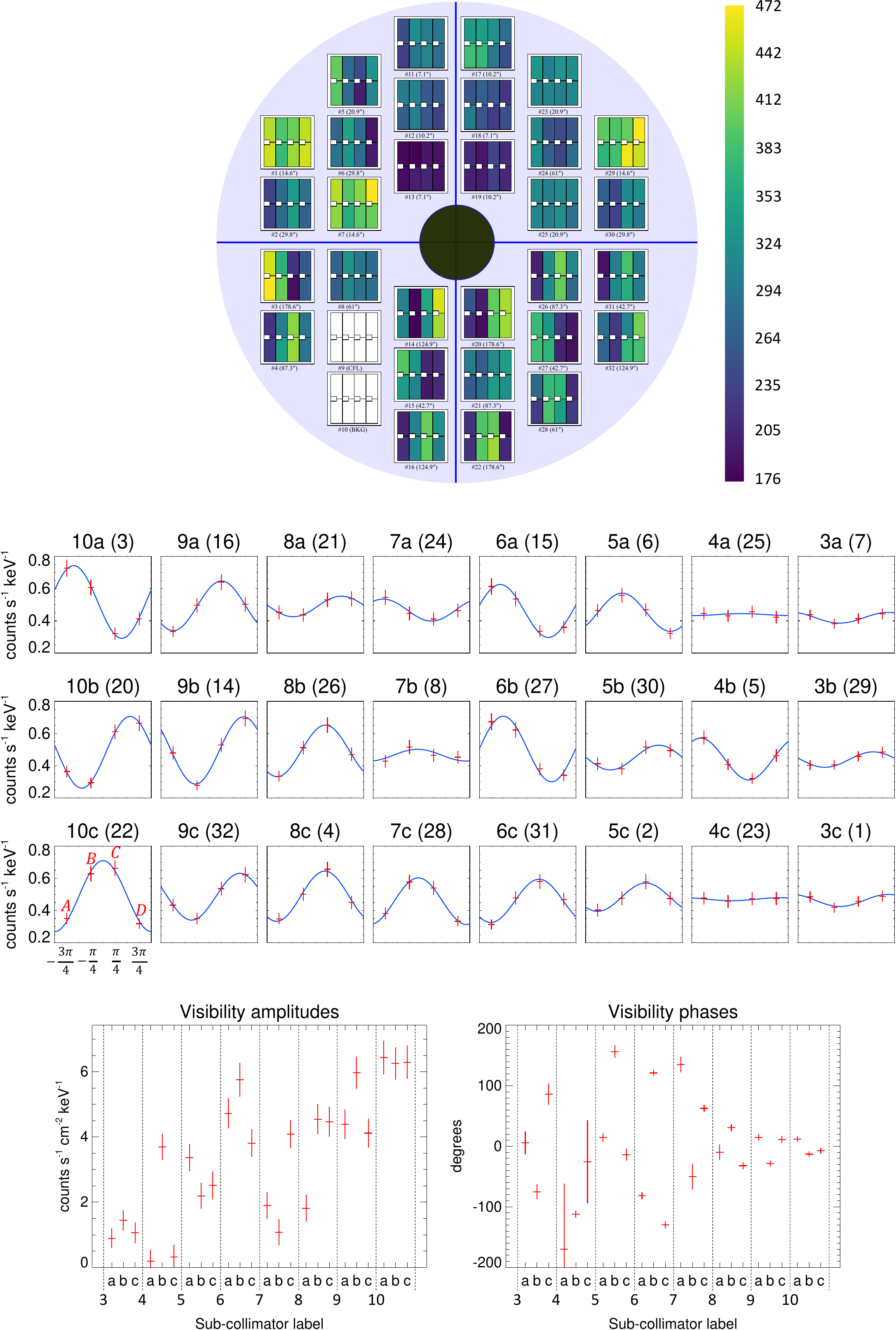}
\caption{\textit{Top panel:} raw counts measured by the STIX pixels during the SOL2022-03-31T18 event, in the time interval 18:26:20-18:27:00 UT and in the energy range 25–50 keV. 
\textit{Middle panel:} background-subtracted count rates corresponding to the pixels of the 24 coarsest STIX sub--collimators. In each panel, the count rate values $A$, $B$, $C$, and $D$ and the associated $3\sigma$ error bars are plotted in red, while the function $g(\beta)$ fitting the count rates (Eq. \eqref{eq:g_beta}) is reported in blue.
\textit{Bottom panels:} visibility amplitude and phase values (left and right panel, respectively), and associated $1\sigma$ error bars, obtained from the experimental count rates measured by the 24 coarsest STIX sub--collimators.}
\label{fig:counts}
\end{figure*}

\subsection{MEM$\_$GE}
This maximum entropy code \citep{massa2020mem_ge} realizes a tradeoff between data fitting and regularization by minimizing the sum of the $\chi^2$ with respect to the observations and the negative entropy functional.
Essentially, the method selects the smoothest solution, which is the one with maximum entropy, among those that fit the data with the same accuracy.
Hence, the selected solution does not present spurious features given by over--fitting of noisy data.
The algorithm applies also two constraints to the solution, in order that the entries are non--negative and the total flux (i.e., the sum of the pixel values) is equal to an a priori estimate.

\subsection{VIS\_FWDFIT based on Particle Swarm Optimization}

VIS\_FWDFIT \citep{2022A&A...668A.145V} returns the values of the parameters associated with a parametric configuration by forward--fitting experimental visibilities.
The parametric configuration is pre--selected by the user among a Gaussian circular source, a Gaussian elliptical source, a \textit{loop} source, or a combination of them \citep[see Fig. 1 of][]{2022A&A...668A.145V}.
Parameter optimization is performed by means of the Particle Swarm Optimization algorithm \citep[PSO;][]{eberhart1995particle}, which is a stochastic method particularly robust to the presence of local minima in the objective function.

\subsection{The SOL2022-03-31T18 event}

\begin{figure*}[ht]
\centering
\includegraphics[width=0.9\textwidth]{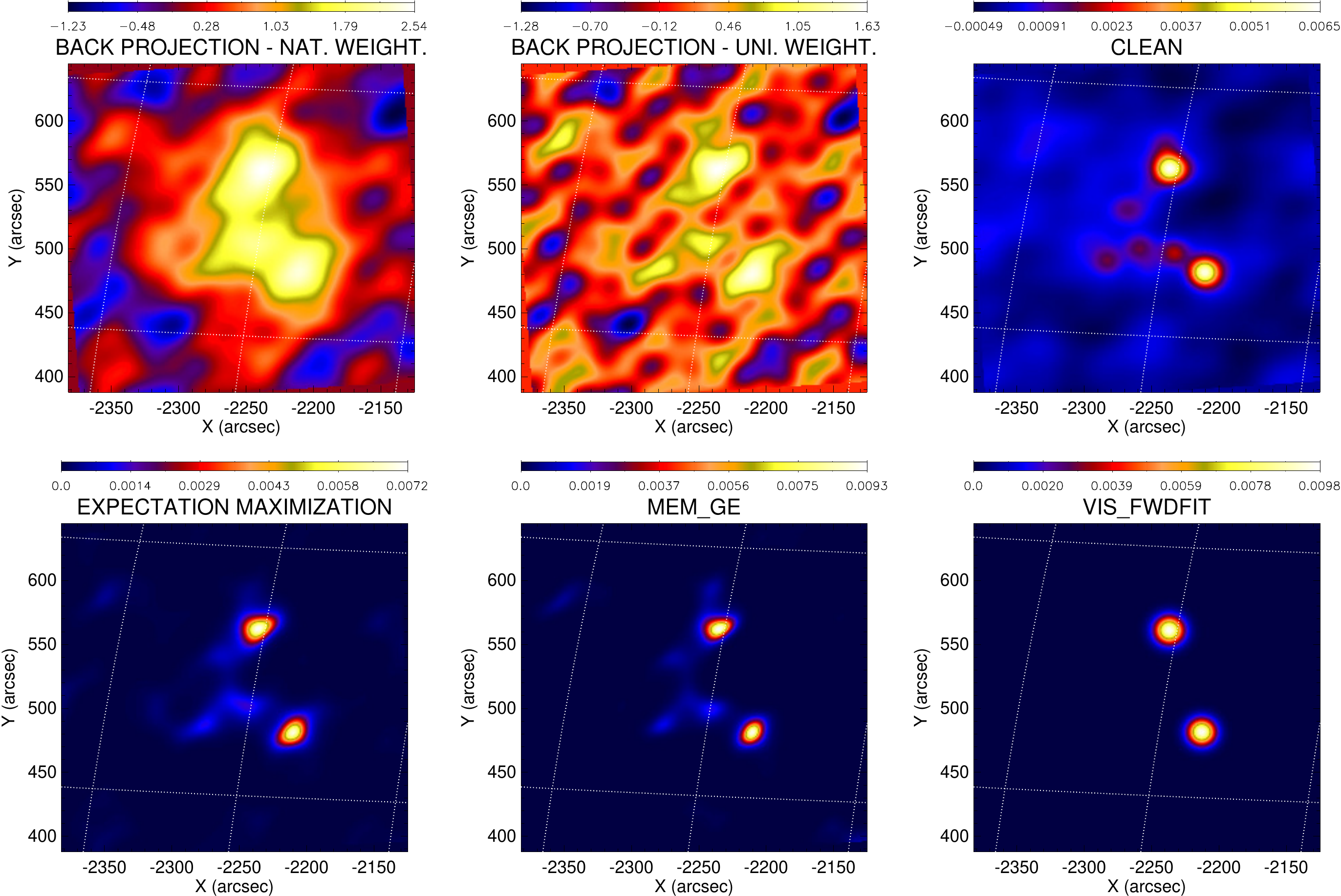}
\caption{Reconstructions of the SOL2022-03-31T18 event in the time interval 18:26:20--18:27:00 UT and in the energy interval 25--50 keV.
\textit{Top row:} reconstructions obtained with Back Projection with natural weighting,Back Projection with uniform weighting and CLEAN, from left to right, respectively.
\textit{Bottom row:} reconstructions obtained with Expectation Maximization, MEM\_GE and VIS\_FWDFIT based on PSO, from left to right, respectively.
Note that a roto--translation has been applied to the maps, so that they are now conceived in the N-W coordinate frame defined on the solar disk and not in the $(x,y)$ reference frame defined in Sect. \ref{section:assumptions} (see also Fig. \ref{fig:reference_frames}).}
\label{fig:reconstructions}
\end{figure*}

As an example of the performances of the different imaging techniques, we compare the results obtained by applying the methods described above in the case of the SOL2022-03-31T18 event integrated in the time range 18:26:20-18:27:00 UT and in the energy range 25--50 keV.
A total of \(\sim \)55,200 counts have been recorded by STIX within these time and energy intervals, \(\sim \)7,600 of which are estimated to be due to background emission.
According to our experience, this is a case with very high statistics. Indeed, as shown by \cite{ZoStiefel2023}, it is possible to reconstruct a rather complex configuration consisting of four sources at similar intensities already with \(\sim \)10,000 counts.

The process that, starting from the raw counts measured by STIX during the considered event, leads to the derivation of the visibility amplitude and phase values is shown in Fig. \ref{fig:counts}. 
The detector location within the DEM (as seen from the Sun side) is reported in the top panel, and the values of the raw counts measured by each detector pixel are indicated with a color table. 
Note that the counts corresponding to the small pixels and to the BKG and CFL sub--collimators are not reported.
The raw counts are then background subtracted and normalized by the detector live time and by the width of the energy interval considered, to obtain calibrated count rates\footnote{To calibrate the count rates (by accounting for effects, such as the the energy dependence of the grid response and internal shadowing of the grids) additional corrections are performed, which will be described in the second paper of the series.}. The values of the count rates and of the corresponding $3\sigma$ uncertainties are reported in red in the middle panel of Fig. \ref{fig:counts}, while the fitting function $g(\beta)$ (defined in Eq. \eqref{eq:g_beta}) is plotted in blue.
The fitting procedure, which has an analytical solution given by Eq. \eqref{eq:vis_amp_phase}, provides the values of the visibility amplitudes and phases. 
These values, and the corresponding $1\sigma$ error bars (including 5\% systematic errors added in quadrature), are shown in the bottom panels of Fig. \ref{fig:counts}. 
Since the current data calibration is satisfactory only for the 24 coarsest sub--collimators \citep{massa2022hard}, we discarded the data recorded by the six finest sub--collimators (labelled with $1$-$a,b,c$ and $2$-$a,b,c$) for the computation of the count rates and of the visibility values.
Note that the visibility phases measured by the coarsest sub--collimators are close to zero since a phase factor is added in order to reconstruct the flaring source at the center of the image.

The reconstructions obtained by applying the different imaging methods are illustrated in Fig. \ref{fig:reconstructions}.
For the Back Projection method, we performed the reconstructions considering both the natural and the uniform weighting for the discretization of the Fourier integral. 
Further, for the CLEAN method we started from a back-projected image obtained with uniform weighting and
we selected a CLEAN beam width of 16.5 arcsec (see Sect. 3.1 of \citet{massa2022hard} for a discussion on the selection of the CLEAN beam width).
In Fig. \ref{fig:reconstructions} we reported the CLEAN reconstruction with added residuals.
Finally, for the VIS\_FWDFIT method we selected a double Gaussian circular source configuration, which is appropriate in the case of a double footpoint configuration.

The different imaging methods provide similar reconstructions (consistent with a non--thermal double footpoint configuration), in terms of number of sources, and their dimension, orientation and separation.  
The only exceptions are represented by the Back Projection reconstructions obtained with both natural and uniform weighting.
Indeed, in the case of natural weighting, the two sources are clearly identified by the method, although they are not fully resolved.
In the case of uniform weighting, instead, the two sources are resolved, but their dimension is larger compared to that of the reconstructions provided by EM, CLEAN, MEM\_GE and VIS\_FWDFIT.
The Back Projection reconstructions are also highly affected by ringing artifacts (especially in the case of uniform weighting, which enhances the high frequency content) due to the limited number of Fourier components sampled by STIX and, hence, available for image reconstructions.
However, this is a known issue of the Back Projection method \citep[see, e.g.,][]{pianabook}, which is a straightforward Fourier inversion and does not provide a regularized solution. 
The other methods return very consistent results, although the sources reconstructed with MEM\_GE are narrower compared to those retrieved by EM, CLEAN and VIS\_FWDFIT.
This is likely due to the fact that the adopted maximum entropy regularization gives rise to super--resolution effects which could lead to slightly narrower reconstructions, as pointed out in simulation tests by \citet{massa2020mem_ge}.

We finally point out that this analysis is not intended as a thorough comparison between the different reconstruction methods, in order to discuss their characteristics and to highlight their specific strengths and limitations when applied to STIX data.
Instead, it can be interpreted as an example of application of the imaging techniques to the same dataset and as a demonstration of the imaging capabilities of the STIX instrument based on the state--of--the--art data calibration \citep{massa2022hard}.
Validation and comparison of the algorithms implemented for image reconstruction from STIX data will be material of future work.

\section{Conclusions}\label{section:conclusions}

In this paper, the first of a series of 3 works on STIX imaging, we described the imaging concept of the instrument.
Specifically, we provided a description of the
mathematical model of STIX data,
under the hypothesis that the considered instrument is \textit{ideal}.
The model we derived can be extended to the case of a \textit{real} instrument by applying appropriate corrections during the data calibration process.
These corrections and, more in general, the STIX data calibration, will be discussed in the second and in the third paper of this series. 
Hence, this work represents a fundamental starting point for the description, calibration and analysis of STIX data.

We showed that each STIX sub--collimator measures a specific Fourier component (or visibility) of the angular distribution of the flaring X--ray source.
In particular, we characterized the angular frequencies sampled by STIX based on the hardware parameters of the sub--collimators, and we showed how the observed counts in an individual detector can be used to directly 
compute the amplitude and phase of the STIX visibility corresponding to that sub-collimator (Eq. \eqref{eq:vis_amp_phase}).
Further, we described the count distribution model for STIX, which is the map that projects the X--ray photon flux into the pixel counts and which allows addressing the imaging problem for STIX directly from the experimental counts. 
We provided an overview of the different FOVs of the instrument which are determined by the choice of the detector pixels that are considered for image reconstruction.
Finally, we defined the image reconstruction problem for STIX based on the derived visibility formation process and on the count distribution model, and we briefly described the implemented imaging methods.
We concluded the work by showing an example of application of the imaging algorithms to experimental STIX data recorded during the SOL2022-03-31T18 flaring event. 

Hard X-ray imaging spectroscopy is a fundamental tool for the understanding of solar flare physics. 
Imaging--spectroscopy from STIX data allows the determination of the angular distribution of the flaring X--ray source at different energies, thus providing crucial information on the underlying electron acceleration mechanisms.
Further, by solving a regularized bremsstrahlung inversion and by applying standard imaging techniques to electron visibilities, it is possible to directly retrieve the images of the electron distribution at different energies, which reveals the actual physics of the X-ray emission \citep[see, e.g.,][]{2007ApJ...665..846P}.
An IDL software for reconstruction of electron maps from STIX data is currently being developed and will be released soon within the STIX Ground Software.
From reconstructed STIX images it is also possible to determine quantitative parameters describing the plasma and the acceleration mechanisms \citep[e.g.,][]{2011SSRv..159..301K}.
Finally, by combining images reconstructed from STIX and HXI data, it will be possible to perform stereoscopic analyses of jointly observed solar flares \citep{Krucker_2019}.
All these studies provide meaningful information only if STIX data are precisely interpreted and calibrated; hence, the work presented in this paper is crucial for the derivation of physically relevant results.

\begin{acknowledgements}
{\em{Solar Orbiter}} is a space mission of international collaboration between ESA and NASA, operated by ESA. The STIX instrument is an international collaboration between Switzerland, Poland, France, Czech Republic, Germany, Austria, Ireland, and Italy. AFB, HC, GH, MK, HX, DFR and SK are supported by the Swiss National Science Foundation Grant 200021L\_189180 and the grant 'Activités Nationales.
PM, EP, FB, AMM, SGU and MP acknowledge the financial contribution from the agreement ASI-INAF n.2018-16-HH.0. 
SGA acknowledges the financial support from the ``Accordo ASI/INAF Solar Orbiter: Supporto scientifico per la realizzazione degli strumenti Metis, SWA/DPU e STIX nelle Fasi D-E''.
SM acknowledges funding from ESA PRODEX via Enterprise Ireland.
FS acknowledges support by the DLR, grant No. 50 OT 1904.

\end{acknowledgements}

\bibliographystyle{aa.bst}
\bibliography{bib_stix}

\begin{appendix}

\section{Derivation of the sub--collimator transmission function $T$}\label{appendix:sub-collimator_transm}

\begin{figure*}[ht]
\centering
\includegraphics[height=4.2cm]{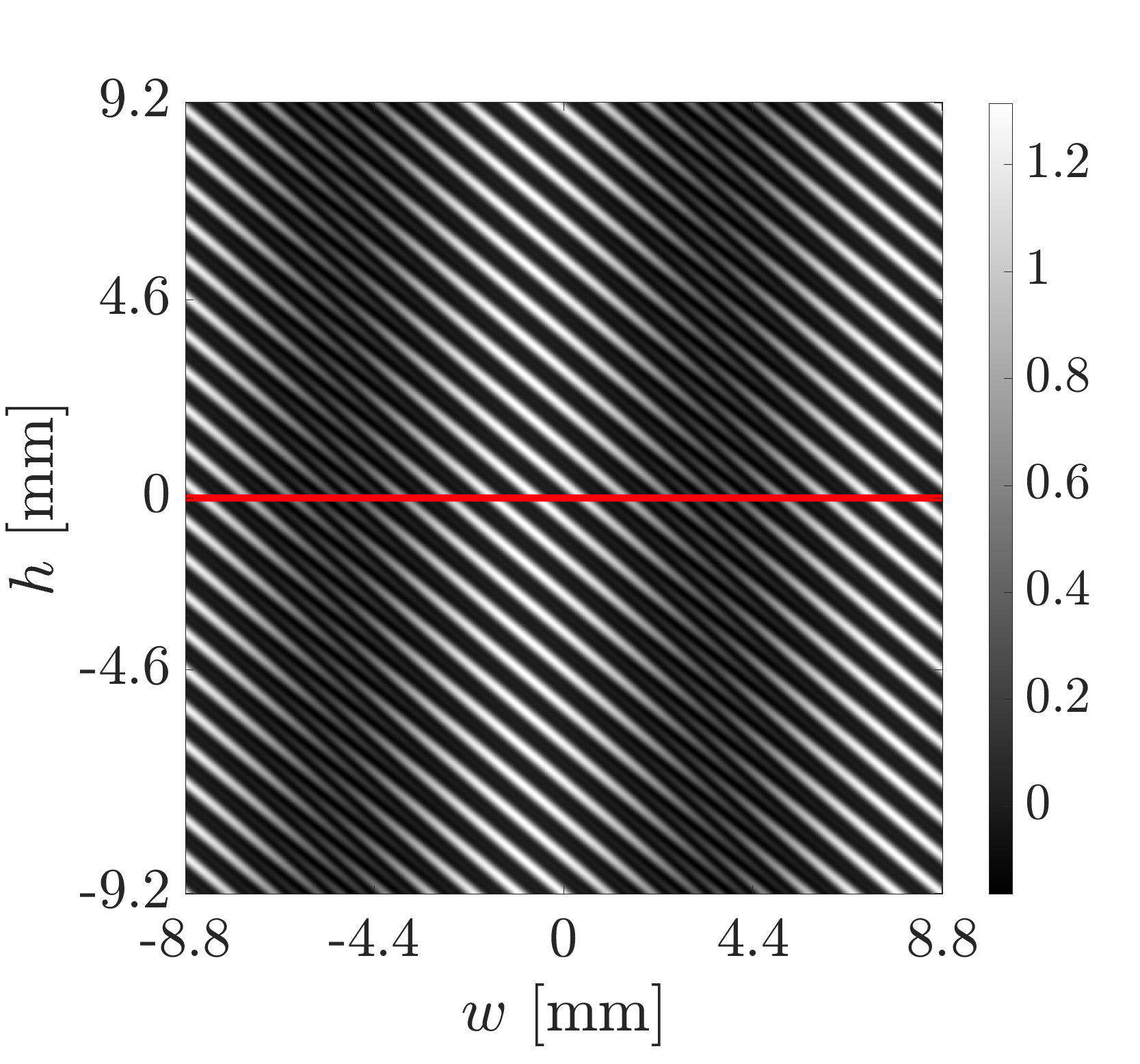}
\hspace{-10pt}
\includegraphics[height=4.2cm]{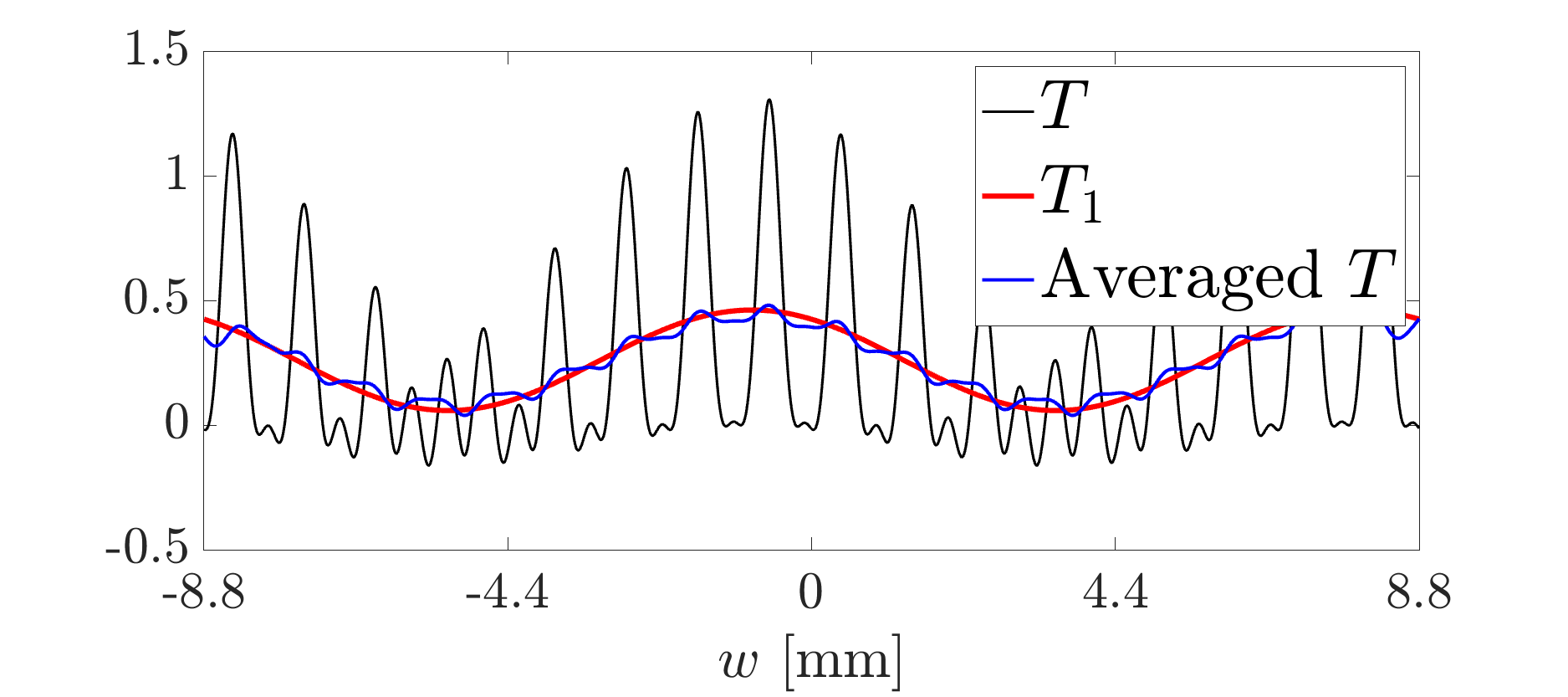}
\caption{\textit{Left panel:} intensity of the transmission function $T$ of sub--collimator 32 ($9c$) computed at $(x,y)=(0,0)$. \textit{Right panel:} profile of $T$ along the red line shown in the left panel (black), profile of the low frequency term $T_1$ along the same line (red), and averaged profile of $T$ obtained by applying a moving mean (blue).
Note that the width of both plots is twice the width of a detector; hence, two periods of the transmission function are plotted.}
\label{fig:moire_profile}
\end{figure*}

In the following, we describe the computations and approximations that lead to the mathematical expression of the sub--collimator transmission function, introduced in Eq. \eqref{eq: transmission}.   
As discussed in the main text of the paper, the transmission functions of the front and rear windows are obtained by replacing $s$, $p$ and $L$ in Eqs. \eqref{eq:w,h,x,y} and \eqref{eq:transm_window} with $s_f$, $p_f$, $d_{\text{sep}}+d_{\text{det}}$ and with $s_r$, $p_r$, $d_{\text{det}}$, respectively.
Denoting
\begin{equation}\label{eq: zeta_f e zeta_r}
\begin{split}
\zeta_f &\coloneqq (w + (d_{\text{sep}}+d_{\text{det}})x, h + (d_{\text{sep}}+d_{\text{det}})y) \cdot \mathbf{k}_f ~,\\
\zeta_r &\coloneqq (w + d_{\text{det}}x, h + d_{\text{det}}y) \cdot \mathbf{k}_r ~,
\end{split}
\end{equation}
the transmission functions are given by
\begin{equation}
\begin{split}
\mathcal{T}_f(w,h,x,y) &\coloneqq \frac{s_f}{p_f} + \frac{2}{\pi} \sin\left( \frac{\pi s_f}{p_f} \right)\cos \left( 2\pi \frac{\zeta_f - \varphi_f}{p_f} \right) ~,\\
\mathcal{T}_r(w,h,x,y) &\coloneqq \frac{s_r}{p_r} + \frac{2}{\pi} \sin\left( \frac{\pi s_r}{p_r} \right) \cos \left( 2\pi \frac{\zeta_r -\varphi_r}{p_r} \right)~.
\end{split}
\end{equation}
Therefore, the overall transmission of a sub--collimator, which is the product of the transmissions of both windows, is
\begin{equation}\label{eq: transmission_appendix}
\begin{split}
T&(w,h,x,y) = \mathcal{T}_f(w,h,x,y) \, \mathcal{T}_r(w,h,x,y) =\\
=& \frac{s_f s_r}{p_f p_r} + \frac{2 s_r}{\pi p_r} \sin \left( \frac{\pi s_f}{p_f} \right) \cos \left( 2\pi \frac{\zeta_f -\varphi_f}{p_f} \right) + \\
&+\frac{2 s_f}{\pi p_f} \sin \left( \frac{\pi s_r}{p_r} \right) \cos \left( 2\pi \frac{\zeta_r - \varphi_r}{p_r} \right) +\frac{4}{\pi^2} \sin \left( \frac{\pi s_f}{p_f} \right) \cdot\\
&\cdot \sin \left( \frac{\pi s_r}{p_r} \right) \cos \left( 2\pi \frac{ \zeta_f - \varphi_f}{p_f} \right)\cos \left( 2\pi \frac{\zeta_r -\varphi_r}{p_r} \right) ~.
\end{split}
\end{equation}
By applying the formula $\cos \alpha \cos\beta = \frac{1}{2}[\cos(\alpha + \beta) + \cos(\alpha - \beta)]$ to the last term of \eqref{eq: transmission_appendix}, we can rewrite the transmission function $T$ as the sum of four terms:
\begin{equation}
\begin{split}
T_1(w,h,x,y) &\coloneqq \frac{s_f s_r}{p_f p_r} +\frac{2}{\pi^2} \sin \left( \frac{\pi s_f}{p_f} \right) \sin \left( \frac{\pi s_r}{p_r} \right) \cdot \\
&\cdot \cos \left( 2\pi \left(\frac{\zeta_f}{p_f} - \frac{\zeta_r}{p_r}\right) - 2\pi\left( \frac{\varphi_f}{p_f} - \frac{\varphi_r}{p_r} \right)\right) ~,\\
T_2(w,h,x,y) &\coloneqq \frac{2 s_r}{\pi p_r} \sin \left( \frac{\pi s_f}{p_f} \right) \cos \left( 2\pi \frac{\zeta_f}{p_f} - 2\pi \frac{\varphi_f}{p_f} \right) ~, \\
T_3(w,h,x,y) &\coloneqq  \frac{2 s_f}{\pi p_f} \sin \left( \frac{\pi s_r}{p_r} \right)\cos \left( 2\pi \frac{\zeta_r}{p_r} - 2\pi \frac{\varphi_r}{p_r} \right) ~, \\
T_4(w,h,x,y) &\coloneqq \frac{2}{\pi^2} \sin \left( \frac{\pi s_f}{p_f} \right) \sin \left( \frac{\pi s_r}{p_r} \right) \cdot \\
&\cdot \cos \left( 2\pi \left(\frac{\zeta_f}{p_f} + \frac{\zeta_r}{p_r}\right) - 2\pi\left( \frac{\varphi_f}{p_f}+\frac{\varphi_r}{p_r} \right)\right) ~.
\end{split}    
\end{equation}
By using Eqs. \eqref{eq: wave vector relations} and \eqref{eq: zeta_f e zeta_r}, and the fact that $\cos(m \alpha) = \cos(\alpha)$ (where $m$ is the $\pm 1$ factor defined at the right--hand side of \eqref{eq: wave vector relations}), the first term becomes
\begin{equation}
\begin{split}
T_1(w,h,x,y) &= \frac{s_f s_r}{p_f p_r} + \frac{2}{\pi^2} \sin \left( \frac{\pi s_f}{p_f} \right) \sin \left( \frac{\pi s_r}{p_r} \right) \cdot \\
&\cdot \cos \left(\frac{2\pi w}{W} + 2\pi (xu+yv) - c_{\varphi} \right) ~,
\end{split}
\end{equation}
where $(u,v)$ and $c_{\varphi}$ are defined as in Eq. \eqref{eq: def (u,v) and c}.
Hence, $T_1$ is a sinusoidal wave with period equal to $W$ and direction parallel to the $w$-axis of the detector coordinate system. 
Moreover, we notice that $T_2$, $T_3$, and $T_4$, if considered as functions of the only variables $(w,h)$, represent high frequency bi-dimensional waves, whose periods are equal to $p_f$, $p_r$, and approximately equal to $\frac{p_f p_r}{p_f + p_r}$, respectively.
Since the pitch of the front and rear windows is much narrower than the dimensions of the detector pixels, the contribution of these high frequency terms averages out when the transmitted X-ray photon flux is integrated over the pixel areas.
Therefore, we can neglect the high frequency terms and approximate the overall sub--collimator transmission as $T \approx T_1$. 
To show that this approximation is accurate, we report in the left panel of Fig. \ref{fig:moire_profile} the intensity of the transmission function $T$ of sub--collimator 32 ($9c$) computed at $(x,y)=(0,0)$.
In the right panel of the same figure, we plot in black the profile of $T$ along the red line parallel to the $w$-axis that is shown in the left panel. 
We overplot in red the profile of $T_1$ along the same line and in blue the average of the profile of $T$ obtained by applying a moving mean with window size equal to the width of a detector pixel.
From the right panel, we can appreciate how $T_1$ represents a good approximation of the averaged profile of $T$, thus demonstrating that the contribution of the high frequencies of the transmission function averages out when the transmitted flux is integrated over a pixel area.

\section{Link between visibility amplitude and phase and pixel counts}\label{appendix:visib_amp_phase}

\begin{figure}[ht]
\centering
\includegraphics[width=0.35\textwidth]{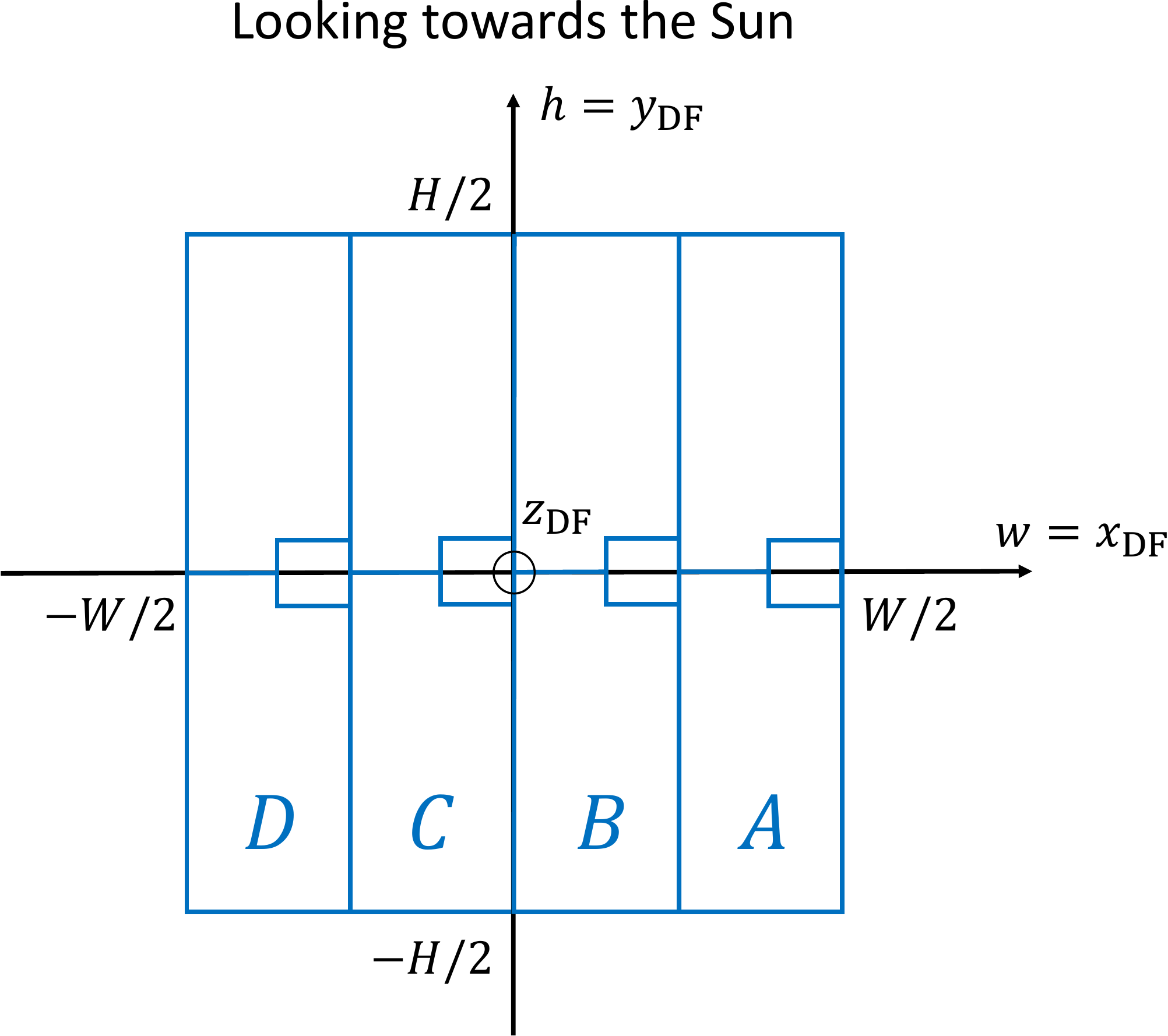}
\caption{Detector reference frame as seen looking from behind the detector towards the Sun. The coordinates of the detector edges are reported in the figure and the number of pixel counts $A$, $B$, $C$, and $D$ recorded by the corresponding pixels are indicated. Compared to the right panel of Fig. \ref{fig:scheme-grid}, the detector appears reversed along the $w$-axis because of the different vantage point. 
Note that the $z_{\text{DF}}$-axis points towards the Sun.}
\label{fig:det_reference_frame}
\end{figure}

We describe below the derivation of Eqs. \eqref{eq: counts ABCD 1} and \eqref{eq:vis_amp_phase}.
We recall that the number of counts recorded by the detector pixels is given by the integral of the Moir\'e pattern over the corresponding pixel area, and that we consider as pixels the four vertical stripes that constitute a detector. 
The number of pixel counts are then 
\begin{equation}\label{eq: counts as int}
\begin{split}
A &= \int_{\frac{W}{4}}^{\frac{W}{2}} \int_{-\frac{H}{2}}^{\frac{H}{2}} M(w,h) \,dw \,dh ~,\\
B &= \int_{0}^{\frac{W}{4}} \int_{-\frac{H}{2}}^{\frac{H}{2}} M(w,h) \,dw \,dh ~, \\
C &= \int_{-\frac{W}{4}}^{0} \int_{-\frac{H}{2}}^{\frac{H}{2}} M(w,h) \,dw \,dh ~,\\
D &= \int_{-\frac{W}{2}}^{-\frac{W}{4}} \int_{-\frac{H}{2}}^{\frac{H}{2}} M(w,h) \,dw \,dh ~. \\
\end{split}
\end{equation}
In Eq. \eqref{eq: counts as int} we considered the detector as seen looking from behind and towards the Sun (see Fig. \ref{fig:det_reference_frame}).
Equation \eqref{eq: counts ABCD 1} can be then obtained by replacing the expression \eqref{eq: express moire} of the Moir\'e pattern in \eqref{eq: counts as int}, and by observing that
\begin{equation}\label{eq:magic_formula}
\begin{split}
&\int_a^b \cos\left(\frac{2\pi w}{W} + \omega - c_{\varphi} \right) \,dw = \\
&= \frac{W}{2\pi} \left[\sin\left(\frac{2\pi b}{W} + \omega -c_{\varphi} \right) - \sin \left(\frac{2\pi a}{W} + \omega -c_{\varphi} \right)\right] = \\
&= \frac{W}{\pi} \cos \left(\frac{\pi}{W}(b+a) + \omega -c_{\varphi} \right) \sin\left(\frac{\pi}{W}(b-a) \right)=\\
&=\frac{W}{\sqrt{2}\pi} \cos \left(\frac{\pi}{W}(b+a) + \omega -c_{\varphi} \right) ~,
\end{split}
\end{equation}
for all $(a,b)\in\left\{ \left(\frac{W}{4},\frac{W}{2}\right), \left(0,\frac{W}{4}\right), \left(-\frac{W}{4},0 \right),  \left(-\frac{W}{2},-\frac{W}{4}\right) \right\}$.
In particular, in the last equalities of \eqref{eq:magic_formula} we have used the fact that $b-a = \frac{\pi}{4}$ for the considered extremes of integration $a$, $b$.

Now, for retrieving the values of $\omega$ and $\mathcal{A}$ from $A$, $B$, $C$, and $D$ (Eq. \eqref{eq:vis_amp_phase}), we need to rewrite \eqref{eq: counts ABCD 1} in a slightly different form. 
Specifically, by using standard trigonometric identities, one can prove that
\begin{equation}\label{eq: counts ABCD 2}
\begin{split}
A &= \mathcal{P} \left[F - \mathcal{E} \, \mathcal{A}\, \cos\left(\omega - c_{\varphi} - \frac{\pi}{4} \right)\right] ~, \\
B &= \mathcal{P} \left[F - \mathcal{E}\, \mathcal{A}\, \sin\left(\omega - c_{\varphi} - \frac{\pi}{4} \right)\right] ~, \\
C &= \mathcal{P} \left[F + \mathcal{E} \, \mathcal{A}\, \cos\left(\omega - c_{\varphi} - \frac{\pi}{4} \right)\right] ~, \\
D &= \mathcal{P} \left[F + \mathcal{E} \, \mathcal{A}\, \sin\left(\omega - c_{\varphi} - \frac{\pi}{4} \right)\right] ~,
\end{split}
\end{equation}
where $F$, $\mathcal{P}$, and $\mathcal{E}$ are defined as in Eqs. \eqref{eq:tot_flux} and \eqref{eq:p and e}, respectively. 
We note that in Eq. \eqref{eq: counts ABCD 2}, neglecting the factor $\mathcal{P}$, the number of counts $A$, $B$, $C$, and $D$ are given by the sum of a constant term (the total flux transmitted by the sub--collimator) and a term related to a sine/cosine function, which is the same in $A$ and $C$, and in $B$ and $D$, but with opposite sign.
By computing the differences between pixel counts, we then obtain
\begin{equation}
\begin{split}
C-A &= 2\,\mathcal{P}\,\mathcal{E} \, \mathcal{A}\, \cos\left(\omega - c_{\varphi} - \frac{\pi}{4} \right)\\
D-B &= 2\, \mathcal{P}\,\mathcal{E} \, \mathcal{A} \, \sin\left(\omega - c_{\varphi} - \frac{\pi}{4} \right) ~;
\end{split}
\end{equation}
from which it is straightforward to derive the values of $\mathcal{A}$ and $\omega$ reported in \eqref{eq:vis_amp_phase}.

\section{STIX FOVs definition}\label{appendix:fovs_definition}

In order to derive the dimension of the different FOVs presented in Sect. \ref{section:def_fov}, we need to determine which pixels are illuminated depending on the location of the flare.
In the following, we will consider an X--ray source located in $(x_c, y_c)$ with respect to the $(x_{\text{STIX}},y_{\text{STIX}})$ reference frame defined in Sect. \ref{section:assumptions}, and we will denote by $W_{\text{win}}$ and $H_{\text{win}}$ the width and the height of the front window, respectively.

\begin{figure*}[ht]
\centering
\includegraphics[width=0.8\textwidth]{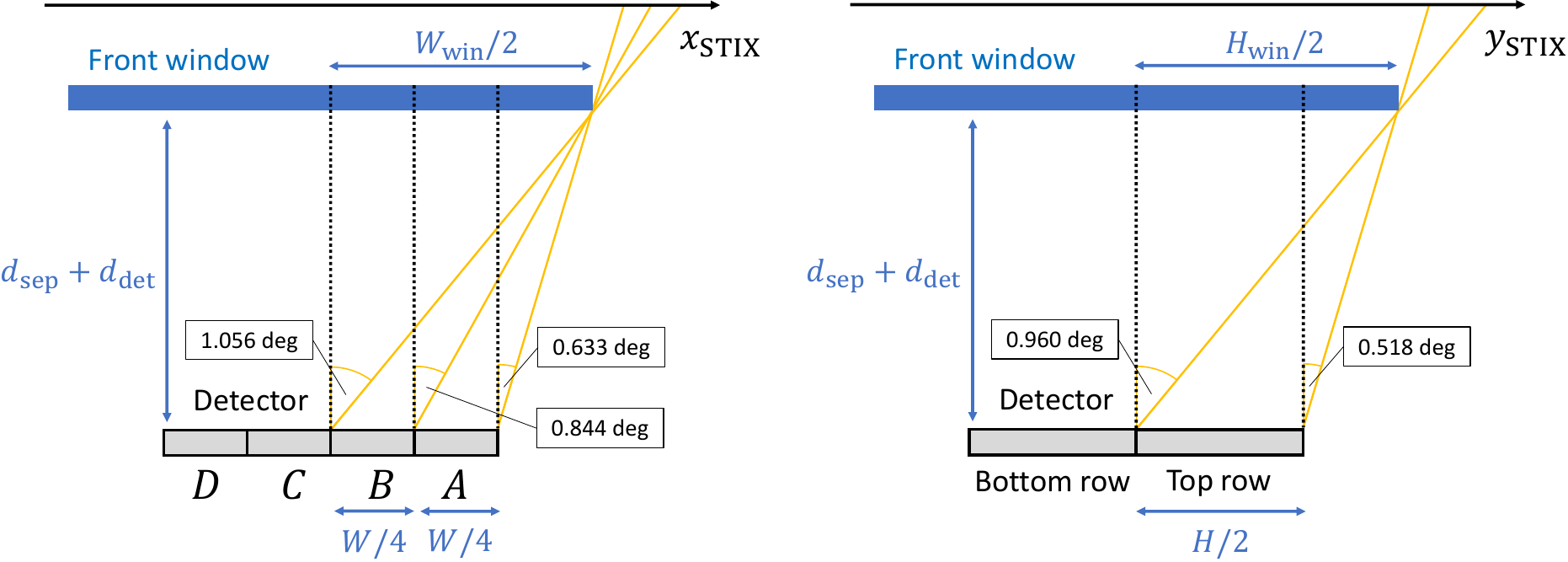}
\caption{Schematics representing the limit angles that define the edges of the different FOVs both in the $x_{\text{STIX}}$ (left panel) and in the $y_{\text{STIX}}$ (right panel) direction.
The profiles of the sub--collimator front window along both the $x$ and the $y$ direction are plotted in blue, while the corresponding profiles of the detector are plotted in gray. The trajectories of the incident photons defining the limit angles are plotted in yellow, and the values of such angles are reported in the panels.}
\label{fig:FOVs_2}
\end{figure*}

We distinguish three cases for the $x_c$ coordinate (see left panel of Fig. \ref{fig:FOVs_2}).

\begin{itemize}
\item The four pixel columns are fully illuminated until one of the vertical edges of the projected rectangular shape of the front window (see the left panel of Fig. \ref{fig:FOVs_1}) coincides with the either the left or the right vertical edge of the detector.
This happens when the $x$ coordinate of the center of the source is 
\begin{equation}\label{eq:x_fov}
\vert x_c \vert = \arctan\left( \frac{W_{\text{win}}/2 - W/2}{d_{\text{sep}}+d_{\text{det}}}  \right) \approx 0.633\,\text{deg} ~.
\end{equation}
Hence, if $\vert x_c \vert < 0.633\,\text{deg}$, then each pixel column is fully illuminated and the corresponding counts can be used for both computing the visibilities and for direct image reconstruction using Eqs. \eqref{eq:vis_amp_phase} and \eqref{eq:count_formation_model}.

\item Three pixel columns are fully illuminated until
\begin{equation}
\vert x_c \vert = \arctan\left( \frac{W_{\text{win}}/2 - W/4}{d_{\text{sep}}+d_{\text{det}}}  \right) \approx 0.844\,\text{deg} ~.
\end{equation}
Therefore, if $0.633\,\text{deg} <\vert x_c \vert < 0.844 \,\text{deg}$, then one of the outer pixel column is only partially illuminated by the X--ray radiation, and the corresponding counts have to be discarded for computing the visibilities or for direct image reconstruction (Eqs. \eqref{eq:vis_amp_phase} and \eqref{eq:count_formation_model}).

\item Two consecutive pixel columns are fully illuminated until
\begin{equation}
\vert x_c \vert = \arctan\left( \frac{W_{\text{win}}/2}{d_{\text{sep}}+d_{\text{det}}}  \right) \approx 1.056\,\text{deg} ~.
\end{equation}
If $0.844\,\text{deg}<\vert x_c \vert <  1.056\,\text{deg}$, we can still compute the values of amplitude and phase of the visibilities using just $A$ and $B$ or $C$ and $D$, provided that an estimate of the total emitted flux $F$ is available (see Equation \eqref{eq: counts ABCD 2}).
Having two adjacent fully illuminated pixel stripes is the minimum requirement for imaging purposes, although some imaging insights can be gained considering just a single pixel per row. 
\end{itemize}

Similar considerations can be made for the $y$ coordinate of the center of the flaring source, although, in this case, we distinguish only two cases (see the right panel of Fig. \ref{fig:FOVs_2}).
\begin{itemize}
\item Top and bottom row pixels are fully illuminated until one of the horizontal edges of the projected rectangular shape of the front window (see the left panel of Fig. \ref{fig:FOVs_1}) coincides with either the top or the bottom horizontal edge of the detector, viz. when
\begin{equation}
\vert y_c \vert = \arctan \left( \frac{
H_{\text{win}}/2 - H/2}{d_{\text{sep}}+d_{\text{det}}} \right) \approx 0.518\,\text{deg} ~.
\end{equation}
In this case, we can use both top and bottom row pixel counts for imaging purposes.
\item Only the top or the bottom row pixels are fully illuminated until
\begin{equation}
\vert y_c \vert = \arctan \left( \frac{H_{\text{win}}/2}{d_{\text{sep}}+d_{\text{det}}} \right) \approx 0.960\,\text{deg} ~.
\end{equation}
If $0.518\,\text{deg}<\vert y_c \vert < 0.960\,\text{deg}$, then we must consider only the counts recorded by the fully illuminated row of pixels when we compute the visibilities using Eq. \eqref{eq:vis_amp_phase} or when we address the image reconstruction problem directly from pixel counts using \eqref{eq:count_formation_model}. 
\end{itemize}

What discussed above is valid for each imaging sub--collimator since the results depend only on the dimensions of the front windows, on the dimensions of the detectors, and on the distance between the front grid and the detectors, and these quantities are the same in each sub--collimator.

\end{appendix}
\end{document}